\documentclass[journal=jctc,manuscript=article]{achemso}
\usepackage[version=3]{mhchem}
\setkeys{acs}{articletitle = true}
\setkeys{acs}{maxauthors=10,etalmode=truncate}
\usepackage{chemformula} 
\usepackage[T1]{fontenc} 
\usepackage{enumitem}
\usepackage{graphicx}
\usepackage{titlecaps}
\Addlcwords{a an the of on at with for and in into from by without}
\usepackage{amsmath}
\usepackage{dcolumn}
\usepackage{bm}
\usepackage{color}
\usepackage{booktabs}
\usepackage[utf8]{inputenc}
\usepackage{mathptmx}
\usepackage{xr}
\SectionNumbersOn

\author{Luis Itza Vazquez-Salazar} \affiliation{Department of
  Chemistry, University of Basel, Klingelbergstrasse 80 , CH-4056
  Basel, Switzerland}

\author{Eric Boittier} \affiliation{Department of Chemistry,
  University of Basel, Klingelbergstrasse 80 , CH-4056 Basel,
  Switzerland}

\author{Oliver T. Unke} \affiliation{Machine Learning Group,
  Technische Universit\"at Berlin, 10587 Berlin, Germany}
\alsoaffiliation{DFG Cluster of Excellence ``Unifying Systems in
  Catalysis'' (UniSysCat), Technische Universit\"at Berlin, 10623
  Berlin, Germany}

\author{Markus Meuwly} \affiliation{Department of Chemistry,
  University of Basel, Klingelbergstrasse 80 , CH-4056 Basel,
  Switzerland.}\alsoaffiliation{Department of Chemistry, Brown
  University, Providence RI, USA} \email{m.meuwly@unibas.ch}

\title{Impact of the characteristics of quantum chemical databases on
  machine learning predictions of tautomerization energies.}

\begin{document}

\begin{abstract}
An essential aspect for adequate predictions of chemical properties by
machine learning models is the database used for training
them. However, studies that analyze how the content and structure of
the databases used for training impact the prediction quality are
scarce. In this work, we analyze and quantify the relationships
learned by a machine learning model (Neural Network) trained on five
different reference databases (QM9, PC9, ANI-1E, ANI-1 and ANI-1x) to
predict tautomerization energies from molecules in Tautobase. For
this, characteristics such as the number of heavy atoms in a molecule,
number of atoms of a given element, bond composition, or initial
geometry on the quality of the predictions are considered. The results
indicate that training on a chemically diverse database is crucial for
obtaining good results but also that conformational sampling can
partly compensate for limited coverage of chemical diversity. The
overall best performing reference database (ANI-1x) performs on
average by 1 kcal/mol better than PC9 which, however, contains about
two orders of magnitude fewer reference structures. On the other hand,
PC9 is chemically more diverse by a factor of $\sim 5$ as quantified
by the number of amons it contains compared with the ANI family of
databases. We explicitly demonstrate that when certain types of bonds
need to be covered in the target database (Tautobase) but are
undersampled in the reference databases the resulting predictions are
poor. Examples include C(sp$^2$)-C(sp$^2$) double bonds close to
hetero atoms and azoles containing N-N and N-O bonds. A quantitative
measure for these deficiencies is the Kullback-Leibler divergence
between reference and target distributions. Analysis of the results
with a TreeMAP algorithm provides deeper understanding of specific
deficiencies in the reference data sets. Capitalizing on this
information can be used to either improve existing databases or to
generate new databases of sufficient diversity for a range of ML
applications in chemistry.
\end{abstract}

\maketitle

\date{\today}

\section{Introduction}
In the last decade, the application of machine learning (ML)
techniques in chemistry has significantly
increased\cite{butler2018machine,unke2020machine,von2020retrospective,noe2020machine}. This
has occasionally been related to a paradigm shift, revolutionizing the
available techniques to understand and simulate
chemistry\cite{agrawal2016perspective, aspuru2018matter}. The
excitement is seemingly justified, given the outcomes of ML
techniques' central promise that, by using a sufficiently large number
of examples and a rule-discovery algorithm, it is possible to obtain a
scientific understanding of the underlying relationships covered by
the data\cite{von2020retrospective, butler2018machine}. Furthermore,
ML techniques are fast compared with quantum chemical methods, while
also reaching comparable
accuracy\cite{behler2007generalized,bartok2010gaussian,rupp2012fast,montavon2013machine,faber2017prediction,schutt2018schnet,unke2018reactive,wilkins2019accurate,veit2020predicting,unke2020high,ko2021fourth}.\\

\noindent
On the other hand, application of quantum ML methods to concrete
problems requires large amounts of data which first need to be
generated from electronic structure
calculations\cite{von2018quantum,heinen2020machine,kaeser2020machine}.
Consequently, data generation is computationally demanding. An
essential challenge for the extension of ML methods' applicability in
chemistry is understanding how suitable databases can be constructed
to maximize accuracy and transferability of the models. An important
ingredient for this step is the degree and confidence with which a
human can understand the relationship between cause (starting database
and model) and result or observation (applying the model to a new
task)\cite{du2019techniques,samek2019towards}. This process has also
been called ``interpretability'' and it can be used to understand the
relationships learned by the model or contained in the data used for
training
it\cite{murdoch2019definitions,dybowski2020interpretable}. Part of the
present work is concerned with the aim to relate the composition of
the initial chemical databases based on which ML models are conceived
with their performance on the prediction of a property of interest
(tautomerization energy) on a set of unseen examples. \\

\begin{figure}[h!]
    \centering
    \includegraphics[width=0.9\textwidth]{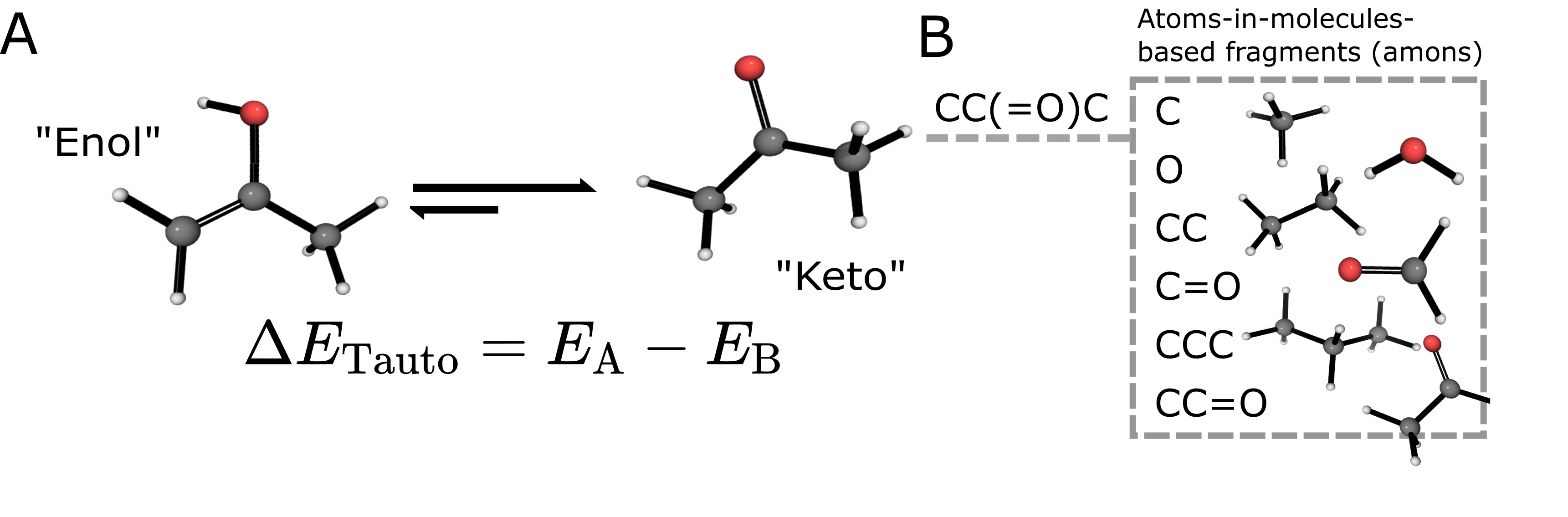}
    \caption{(A) Tautomerism is a form of reversible isomerization
      involving the rearrangement of a charged leaving group within a
      molecule.  The keto-enol tautomerism of acetone, an equilibrium
      which heavily favours the keto side, is shown as an example.
      (B) Chemical space can be decomposed systematically through the
      use of atoms-in-molecules-based fragments
      (amons).\cite{huang2020quantum} The amons present in acetone
      (SMILES: CC(=O)C), as well as their corresponding SMILES are
      given as an example.}
    \label{fig:veen}
\end{figure}

\noindent
To test the effect of different databases on the reliability of the ML
model, the problem of predicting tautomerization energies is
considered.  Tautomerism is a form of reversible isomerization
involving the rearrangement of a charged leaving group within a
molecule\cite{wilkinson1997iupac} (e.g. Fig. \ref{fig:veen}a).  One
isomer transforms into the other by a heterolytic splitting followed
by a recombination of the fragments
formed\cite{raczynska2005tautomeric}.  This process involves the
migration of one or more double bonds and atoms or groups.  The
isomers (i.e.\ tautomers) generated in this reaction are chemically
independent species with defined properties\cite{martin2009let}.  It
is known that this type of reaction is of importance for biological
molecules such as amino acids\cite{raczynska2005tautomeric},
DNA\cite{watson1953molecular,shukla2013tautomerism},
RNA\cite{singh2015role}, and atmospheric
processes.\cite{kaser2020isomerization} Additionally, it is estimated
that tautomerism can occur in up to two thirds of small
molecules\cite{sitzmann2010tautomerism}, and a majority of commercial
drugs\cite{greenwood2010towards,pospisil2003tautomerism}.  \\

\noindent
Despite its widespread occurrence and importance, quantitative studies
of tautomerism are still challenging because small changes in
molecular structure or solvent environment can dramatically change the
tautomeric
equilibrium\cite{martin2009let,taylor2014tautomerism}. Moreover, small
free energy differences between two tautomers in solution make the use
of high level theoretical methods and an adequate basis set mandatory
which limits its use for calculations of tautomerization energies and
ratios. \cite{taylor2014tautomerism,fogarasi2010studies} As an
example, tautomerization in malonaldehyde (MA) is considered. MA has
served as a prime example to develop and test computational methods
for a realistic description of hydrogen transfer in small
molecules.\cite{kaser2020reactive} Experimentally, the ground state
tunneling splitting is 21.58314 cm$^{-1}$ which has been determined by
different experiments with very high
accuracy.~\cite{mawilson81,mafirth91} Furthermore, proton transfer
rates in a di-imine derivative have been determined with nuclear
magnetic resonance (NMR) spectroscopy.~\cite{limbach87} Such
experiments provide direct information on the barrier height
separating the two tautomeric states ``A'' and ``B''. Using a
state-of-the art full-dimensional potential energy surface at the near
basis-set-limit frozen-core CCSD(T) level of theory,\cite{bowman:2008}
the tunneling splitting from quantum simulations was determined as
23.4 cm$^{-1}$.\cite{meyer:2011} Alternatively, using a reduced
dimensionality Hamiltonian, the barrier height for proton transfer in
a parametrized molecular mechanics with proton transfer (MMPT)
potential was found to be 4.34 kcal/mol which yields a tunneling
splitting of 21.2 cm$^{-1}$, consistent with
experiment.\cite{MM.ma:2010,MM.kinetic:2017} This barrier height is
close to the value from CCSD(T) calculations which yield 4.1
kcal/mol.\cite{bowman:2008} These examples illustrate that
calculations at the highest levels of theory are required for
quantitative studies of the energetics underlying tautomerization.\\

\noindent
In the last decade, development of ML models has allowed the design of
robust models that can routinely reach prediction errors better than
chemical accuracy at low computational
cost\cite{schutt2018schnet,unke2019physnet}. However, there have been few discussions on how databases
can be improved/designed to obtain better predictions from the ML
model. Ideally, the combination of a robust ML model and an adequate
database will result in quantitative results for the prediction of a
property of interest.  The availability of public databases of
tautomers\cite{wahl2020tautobase,dhaked2020tautomer} makes the
prediction of tautomerization energies using these ML models an ideal
test case to study how different training databases influence the
accuracy of ML methods.  \\

\noindent
The present work is structured as follows. First, the methods,
databases, and the analysis performed are introduced. Next, the
results for the tautomerization energy predictions using models
trained on the different tested databases are presented. Additionally,
prediction errors for tautomerization energies are analyzed. The
effect of different characteristics of the training data on predicting
the tautomerization energy and the individual molecules' energy are
evaluated. Finally, the results are discussed and conclusions
regarding the findings and interpretability of broadly conceived and
learned ML models applied to a specific chemical question are drawn.

\section{Methods} \label{sec:methods}
\subsection{Machine Learning}
PhysNet was used for the representation and evaluation of the data
sets, using the hyperparameters from the original
publication\cite{unke2019physnet}. For training, only the nuclear
charges ($Z$), the energies of the molecules ($E$) and their
coordinates ($\mathbf{R}$) were considered. The energies used for
training were those reported by the different databases minus the
atomization energy at the given level of theory. In all cases, a
training, testing and validation split of 8 : 1 : 1 was used. The loss
function was
\begin{equation}
    \mathcal{L}=w_{\rm{E}}|E-E^{\rm{ref}}| + \lambda_{\rm{nh}}
    \mathcal{L}_{\rm{nh}}
    \label{eq1}
\end{equation}
where $E^{\rm{ref}}$ is the reference energy, $w_{\rm{E}}=1$ is the
weighting hyperparameter for the energy, $\lambda_{\rm{nh}}=10^{-2}$
is a regularization hyperparameter and the term
$\mathcal{L}_{\rm{nh}}$ is a non hierachicality regularization
penalty. The loss function (Eq. \ref{eq1}) was minimized using AMSgrad
with a learning rate of $10^{-3}$. Overfitting was prevented by the
use of early stopping, the convergence criteria considered is the
saturation of the validation-loss function\cite{unke2019physnet}.\\

\subsection{Database Selection}
For training the NNs, four widely used databases for benchmarking
predictive models of DFT-based energies are employed, namely
QM9\cite{ramakrishnan2014quantum}, PC9\cite{glavatskikh2019dataset},
ANI-1\cite{smith2017ani} and ANI-1x.\cite{smith2020ani} An additional
database, ANI-1E (where "E" stands for equilibrium) containing only the
equilibrium structures of the ANI-1 database, was generated at the
$\omega$B97x/6-31G(d) level of theory. These databases can be divided
into categories based on the type of geometries they contain. The
datasets consisting solely of equilibrium structures are QM9, PC9 and
ANI-1E, and sample only chemical space. Contrary to that, the ANI-1 and
ANI-1x databases contain equilibrium and non-equilibrium structures
which sample chemical and conformational space. \\

\noindent
{\bf Training sets:} The QM9 data set\cite{ramakrishnan2014quantum}
was generated as a subset of the GDB-17 chemical
universe\cite{ruddigkeit2012enumeration}, consisting of 133885
molecules, containing less than or equal to 9 heavy atoms (either C,
N, O or F). Reference energies were computed at the B3LYP/6-31G(2df,p)
level of theory. For the present work, QM9 was filtered to include
only molecules which passed a geometry consistence
check\cite{ramakrishnan2014quantum} and considering only those
containing carbon, nitrogen or oxygen atoms. The final size of the QM9
training dataset used here consisted of 110426 molecules.\\

\noindent
The PC9 dataset was created as an alternative to QM9 to improve
coverage of chemical space.\cite{glavatskikh2019dataset} It is a
subset of the PubChemQC\cite{nakata2017pubchemqc} and is limited to
molecules with 9 heavy atoms or less ($n_{\rm atoms} \leq 9$). This
database consists of 99234 molecules, calculated at the B3LYP/6-31G(d)
level of theory, and excludes enantiomers, tautomers, isotopes as well
as other specific artifacts in
PubChemQC\cite{glavatskikh2019dataset}. PC9 also contains 5325
molecules with an electronic state different from a singlet which were
removed for the present work. As in the case of QM9, molecules which
contain fluorine were removed. The final size of this dataset was
85875 molecules.\\

\noindent
ANI-1\cite{smith2017ani} consists of 24 million geometries generated
using normal mode sampling from 57462 unique molecules. ANI-1 is a
subset of the GDB-11 chemical universe\cite{fink2005virtual,
  fink2007virtual}. A related dataset, ANI-1x \cite{smith2020ani}, was
created using an active learning\cite{smith2018less} procedure which
reduced the original ANI-1 database to 5 million structures. Starting
from the ANI-1 database\cite{smith2017ani,smith2017ani1} the ANI-1E
dataset was generated and consists only of the corresponding
equilibrium structures. The new ANI-1E database contains 57462
molecules limited to eight heavy atoms (either C, O or N). The
generation of this database is further described below in the
subsection of electronic structure calculations.\\

\noindent
{\bf Tautomerization energy evaluation set:} The performance of the NN
models described above was evaluated on a subset of molecules from
TautoBase\cite{wahl2020tautobase}, a public database of 1680 tautomer
pairs. The Tautobase was filtered to molecules only containing
hydrogen, carbon, nitrogen, or oxygen atoms. The size of the final
test set was 1257 tautomer pairs (2514 molecules). The geometry
generation and structural optimization for these molecules is
described below.\\

\noindent
\subsection{Initial geometry}
\label{subsec:initial_geom}
To investigate the effect of the geometry of the molecules passed to
the NN model on its performance, a second set of geometries for the
tautobase was also evaluated. These geometries were generated from the
SMILES representation using OpenBabel\cite{o2011open} and were
optimised with the MMFF94 force field\cite{halgren1996merck}. \\

\noindent
Additionally, a subset of the test set composed of 34 tautomeric pairs
which were part of the SAMPL2 challenge\cite{geballe2010sampl2} were
considered. Those 34 pairs (68 molecules) were optimized by six
popular general atomistic force fields: CHARMM27\cite{foloppe2000all},
GAFF\cite{wang2004development}, OPLS\cite{jorgensen1988opls},
UFF\cite{rappe1992uff}, Gromos\cite{schmid2011definition}, and
Ghemical\cite{hassinen2001new}. Details on the generation of the
geometries are reported in the SI.\\

\subsection{Electronic Structure Calculations}
{\it Generation of ANI-1E:} Starting from the ANI-1
database\cite{smith2017ani1}, a new data set, ANI-1E, was
generated. From the SMILES strings provided by \citet{smith2017ani1}
initial geometries using OpenBabel\cite{o2011open} were
generated. Subsequently, geometries were optimised using
PM7\cite{stewart2013optimization} implemented in
MOPAC2016\cite{mopac2016}, before a final geometry optimization and
frequency calculation at the $\omega$B97x/6-31G(d) level of theory was
performed using Gaussian09\cite{gaussian}. The final results were
checked to assure they did not contain imaginary frequencies and
therefore correspond to minima on the potential energy surface.\\

\noindent
{\it Tautomerization evaluation set:} The molecules used for the
evaluation of the NN models were generated from the SMILES provided in
Ref.~\citenum{wahl2020tautobase} using the OpenBabel
software\cite{o2011open}. These structures were then optimized at the
same level of theory for each of the databases used for training (QM9:
B3LYP/6-31G(2df,p), PC9: B3LYP/6-31G(d), ANI: $\omega$B97x/6-31G(d))
using Gaussian09\cite{gaussian}. Here, the tautomerization energy
$\Delta E_{\rm Tauto}$ is defined as the energy difference between
tautomers A and B in their optimized structures, see
Figure~\ref{fig:veen}. These optimized geometries were given as input
to the respective, previously trained NN models using the five
reference databases QM9, PC9, ANI-1E, ANI-1 and ANI-1x.\\

\begin{table}[hbtp]
\caption{Overview of the training datasets used in this work. QM9, PC9
  and ANI-1E contain equilibrium structures and can be considered to
  only sample chemical space, whereas ANI-1 and ANI-1x also contain
  non-equilibrium geometries which sample conformational space. The
  number of molecules refers to the total number of data in each
  dataset.$^{a}$Structures generated through normal mode
  sampling. $^{b}$Training set selected using active learning.}
\begin{tabular}{ccccc}
\toprule
\multicolumn{1}{c|}{Database} & Number of Molecules & Level of Theory    & Parent Universe                \\ \midrule
\multicolumn{1}{c|}{QM9}      & 128908              & B3LYP/6-31G(2df,p) & GDB-17                          \\
\multicolumn{1}{c|}{PC9}      & 85870               & B3LYP/6-31G(d)     & PubChemQC                \\
\multicolumn{1}{c|}{ANI-1E}    & 57462               & $\omega$B97x/6-31G(d)     & GDB-11          &                      \\ \midrule
\multicolumn{1}{c|}{ANI-1$^{a}$}    & 24 million                 & $\omega$B97x/6-31G(d)     & GDB-11           \\
\multicolumn{1}{c|}{ANI-1x$^{b}$}   & 5 million                  & $\omega$B97x/6-31G(d)     & ANI-1                \\ \bottomrule
\end{tabular}
\label{tab:tab1}
\end{table}

\subsection{Comparison of structural properties of different databases}
As a way to compare the composition of the different datasets
evaluated in terms of structural properties (e.g. bond lengths), a
Gaussian kernel density estimation\cite{diwekar2015probability} of
their distributions was generated, see Figures
\ref{sifig:dist_bonds_C_9} to \ref{sifig:dist_bonds_N_9p}. The
similarities between the distributions used to train the NN models and
those from the test set of tautomers was quantified by computing the
relative entropy (or Kullback-Leibler (KL)
divergence)\cite{cover2006elements}
\begin{equation}
D(p\parallel q) = \int_{-\infty}^{\infty} p(x) log\left( \frac{p(x)}{q(x)} \right) dx
\label{eq:kl}
\end{equation}
This metric quantifies the overlap between a reference distribution
$p(x)$ and a target distribution $q(x)$. Because the KL divergence is
not symmetric ($D(p \parallel q) \neq D(q \parallel p)$), it is
important to specify which distribution is used as the reference. In
the present work, the Tautobase (target) distribution and the QM9,
PC9, ANI-1E are the reference distributions. The KL divergence allows
to quantify how much information of the reference databases (i.e. QM9,
PC9, ANI-1E) is 'missing' to best cover the information contained in
Tautobase.\\

\subsection{Chemical Space ``Coverage'' from Fragment Analysis}
The `coverage' of chemical space contained in the Tautobase, QM9, PC9
and ANI family of databases was analysed with respect to the
atom-in-molecule-based fragments (amons)\cite{huang2020quantum}. The
amons are generated from the SMILES representation of the molecule,
see Figure \ref{fig:veen}b. This representation is used to construct a
molecular graph from which sub-graphs to a maximum number of atoms
(excluding hydrogen) are generated. All sub-graphs are checked to be
valid and unique. Here, amons up to and including a maximum of five
heavy atoms were generated by an in-house script.

\section{Performance of the NN on the Tautobase}
\subsection{Overall Performance}
\begin{figure}[h!]
    \centering
    \includegraphics[width=0.75\textwidth]{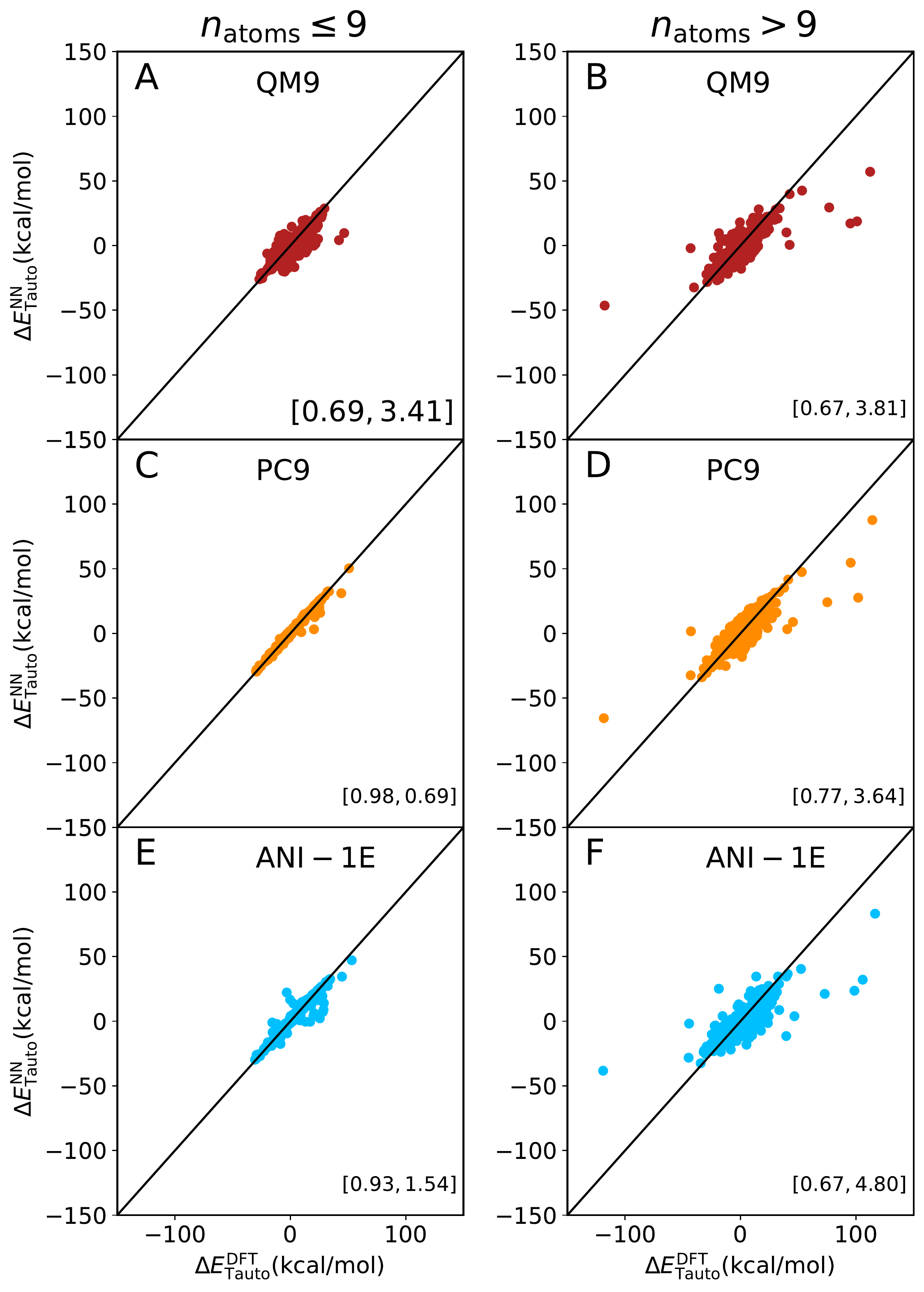}
    \caption{Correlation between the calculated (DFT) and predicted
      (NN) tautomerization energies ($\Delta E_{\rm{Tauto}} =
      E_{\textrm{A}}-E_{\rm{B}}$) for molecules with $n_{\rm{atoms}}
      \leq 9 $ (left) and $n_{\rm{atoms}} > 9 $ (right) for the models
      trained on the datasets which only cover chemical space. Pearson
      correlation coefficients ($r^{2}$) and the Mean Absolute Error
      (MAE) values are reported in brackets as [$r^{2}$, MAE].}
    \label{fig:scatter_chemspace}
\end{figure}

\begin{figure}[h!]
    \centering
    \includegraphics[width=0.75\textwidth]{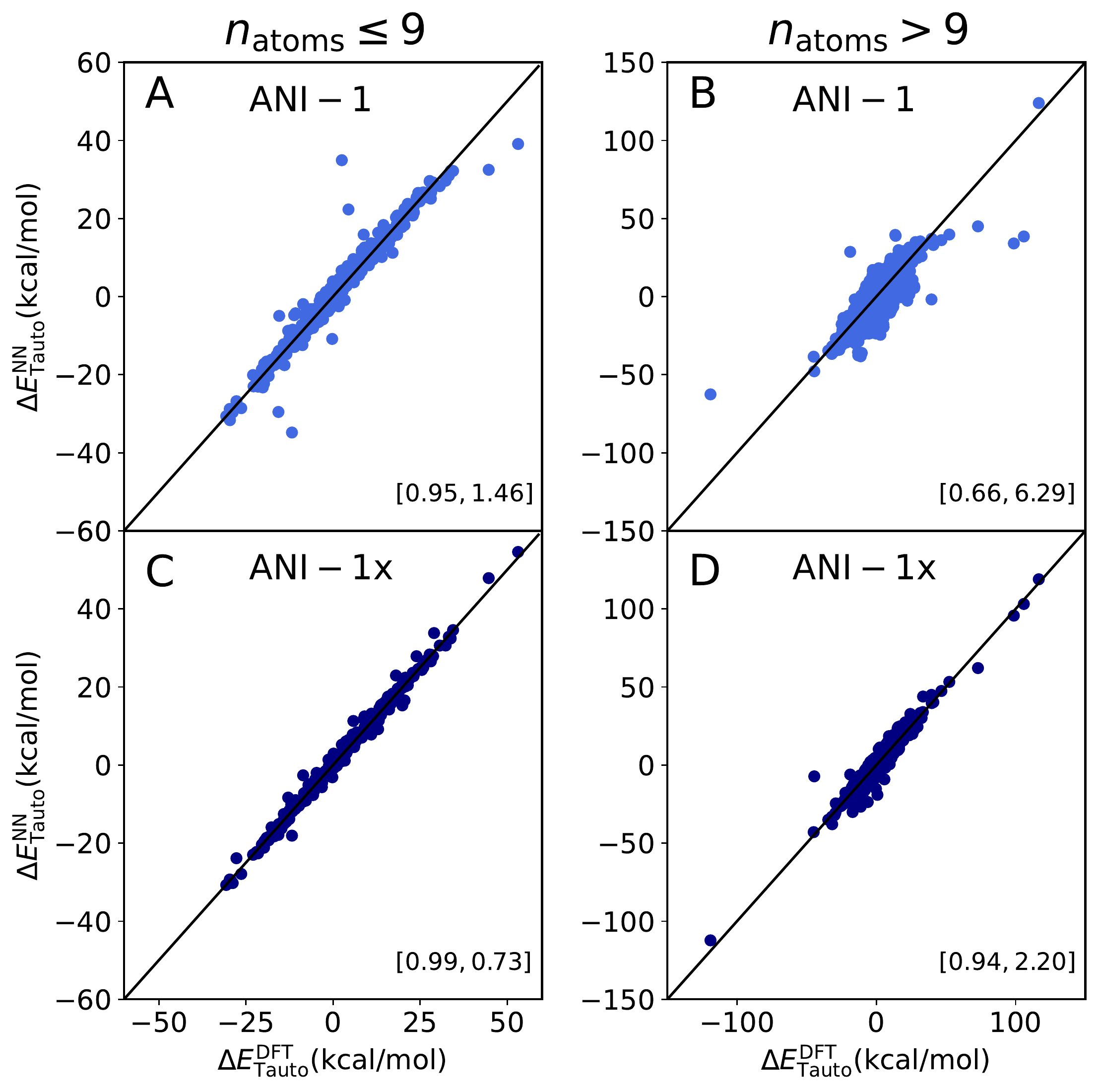}
    \caption{Correlation between the calculated (DFT) and predicted
      (NN) tautomerization energies ($\Delta E_{\rm{Tauto}} =
      E_{\textrm{A}}-E_{\rm{B}}$) for molecules with $n_{\rm{atoms}}
      \leq 9 $ (left) and $n_{\rm{atoms}} > 9 $ (right) for the models
      trained with the datasets which cover chemical and
      conformational space. Pearson correlation coefficients ($r^{2}$)
      and Mean Absolute Error (MAE) values are reported in brackets
      [$r^{2}$, MAE].}
    \label{fig:scatter_chemconfspace}
\end{figure}

The mean absolute errors for the tautomerization energies $\Delta
E_{\rm Tauto}$ range from 1.68 kcal/mol (ANI-1x) to 4.59 kcal/mol
(ANI-1). The results are summarized in Table~\ref{tab:mae} for all
molecules in the test set and graphically reported in
Figures~\ref{fig:scatter_chemspace} (datasets with equilibrium
structures) and ~\ref{fig:scatter_chemconfspace} (datasets with both
equilibrium and non-equilibrium structures).  The prediction errors
for the energy of single isomers, $E_{\rm SI}$, with respect to DFT
energies are also reported (Table \ref{tab:mae}). Note that because
the tautomerization energy is defined as the difference between the
isomer energies, predictions of tautomerization energies are often
more accurate due to cancellation of systematic errors. On the other
hand, the energies for single isomers are considerably larger and span
a much wider range because they scale with the number of atoms that
make up a molecule. Therefore, the NN-based energies for larger
molecules are expected to be associated with considerably larger
errors.\\

\noindent
To assess whether the accuracy for predicting tautomerization energies
correlates with the performance of the trained NNs on the chemical
databases, the QM9, PC9, and ANI-1E models are considered. The MAEs on
held-out test sets for each of the respective training runs are 0.10
kcal/mol (QM9), 0.69 kcal/mol (PC9), and 0.27 kcal/mol (ANI-1E) which
is comparable to values ranging from 0.19 kcal/mol to 0.30 kcal/mol
for the QM9 data set (depending on the sizes of the training and
validation sets used).\cite{unke2019physnet} However, when applying
these trained NNs to evaluate Tautobase, the MAEs are 3.67 kcal/mol
(QM9), 2.60 kcal/mol (PC9), and 3.66 kcal/mol (ANI-1E),
respectively. Hence, there appears to be no correlation between the
quality of the trained NNs (measured on test data sampled from the
same database used for training) and their performance on Tautobase.\\

\begin{table}[hbtp]
\caption{Mean Absolute (MAE) and Root-Mean-Squared Error (RMSE) for
  the prediction of tautomerization energy $\Delta E_{\rm Tauto}$, and
  the single isomer energies, $E_{\rm SI}$, for the entire Tautobase
  (1257 tautomeric pairs) for each of the datasets.}
\begin{tabular}{l|cc|cc|}
\toprule
 & \multicolumn{2}{c}{$ \Delta E_{\textrm{Tauto}}$} & \multicolumn{2}{c}{$E_{\textrm{SI}}$}        \\ 
Database & MAE                 & RMSE               & MAE           & RMSE  \\ \midrule
QM9      & 3.67                & 7.12               & 5.00       &  8.40 \\ 
PC9      & 2.60                & 5.41               & 6.90       &  13.20 \\ 
ANI-1E    & 3.66                & 7.09               & 15.20      & 17.50 \\ 
ANI-1    & 4.59                & 7.56               & 13.40      & 17.00 \\ 
ANI-1x   & 1.68                  & 2.85                 & 1.80       &  3.60  \\ \bottomrule
\end{tabular}
\label{tab:mae}
\end{table}

\noindent
Next, the performance of the trained models for predicting $\Delta
E_{\rm Tauto}$ and $E_{\rm SI}$ depending on the number of heavy atoms
is assessed. For this, results for the subset of molecules with
$n_{\rm atoms} \leq 9$, referred to as ``Set1''  in the following, is considered
separately from those with $n_{\rm atoms} > 9$, which is
``Set2''. This distinction is motivated by the fact that the PC9 and
QM9 databases contain structures with only up to 9 heavy atoms,
i.e.\ models need to extrapolate for larger structures. For Set1, the
PC9 (Figure \ref{fig:scatter_chemspace}C) and ANI-1x (Figure
\ref{fig:scatter_chemconfspace}C) data sets perform best. Both achieve
chemical accuracy (MAE < 1 kcal/mol) with respect to the DFT values
for $\Delta E_{\rm Tauto}$.\\

\noindent
The extrapolation to Set2 increases the prediction errors for most of
the databases studied. Again, the ANI-1x database performed best for
$\Delta E_{\rm{Tauto}}$ with a MAE of 2.20 kcal/mol, followed by PC9,
QM9, ANI-1E and ANI-1 with a MAE of 6.29 kcal/mol. The number of atoms
in the database also influences the correlation coefficient $r^2$. A
better correlation is observed when the size of evaluated molecules is
in the range covered by the training database, i.e. for Set1. Here,
the correlation coefficients range from 0.69 (QM9) to 0.99
(ANI-1x). For Set2, the $r^2$ values are significantly lower. A
particularly noteworthy case is the QM9 database, which shows almost
the same MAE and $r^2$ for both subsets of the tautobase (Figures
\ref{fig:scatter_chemspace}A and B). The performance for Set1 and Set2
also differs in the number and magnitude of outliers, see Figures
\ref{fig:scatter_chemspace} and \ref{fig:scatter_chemconfspace}. \\

\noindent
The RMSE in Table~\ref{tab:mae} show that the spread for $E_{\rm SI}$
could be a reason for large outliers (see Figure
\ref{sifig:error_dist}). It is likely that there is some error
cancellation when predicting tautomerization energies (i.e.\ energy
differences). For example, if the trained NN predicts too large
energies for both isomers, this systematic error cancels when their
energy difference is computed.\\

\subsection{Error analysis}
\begin{figure}
    \centering
    \includegraphics[width=0.85\textwidth]{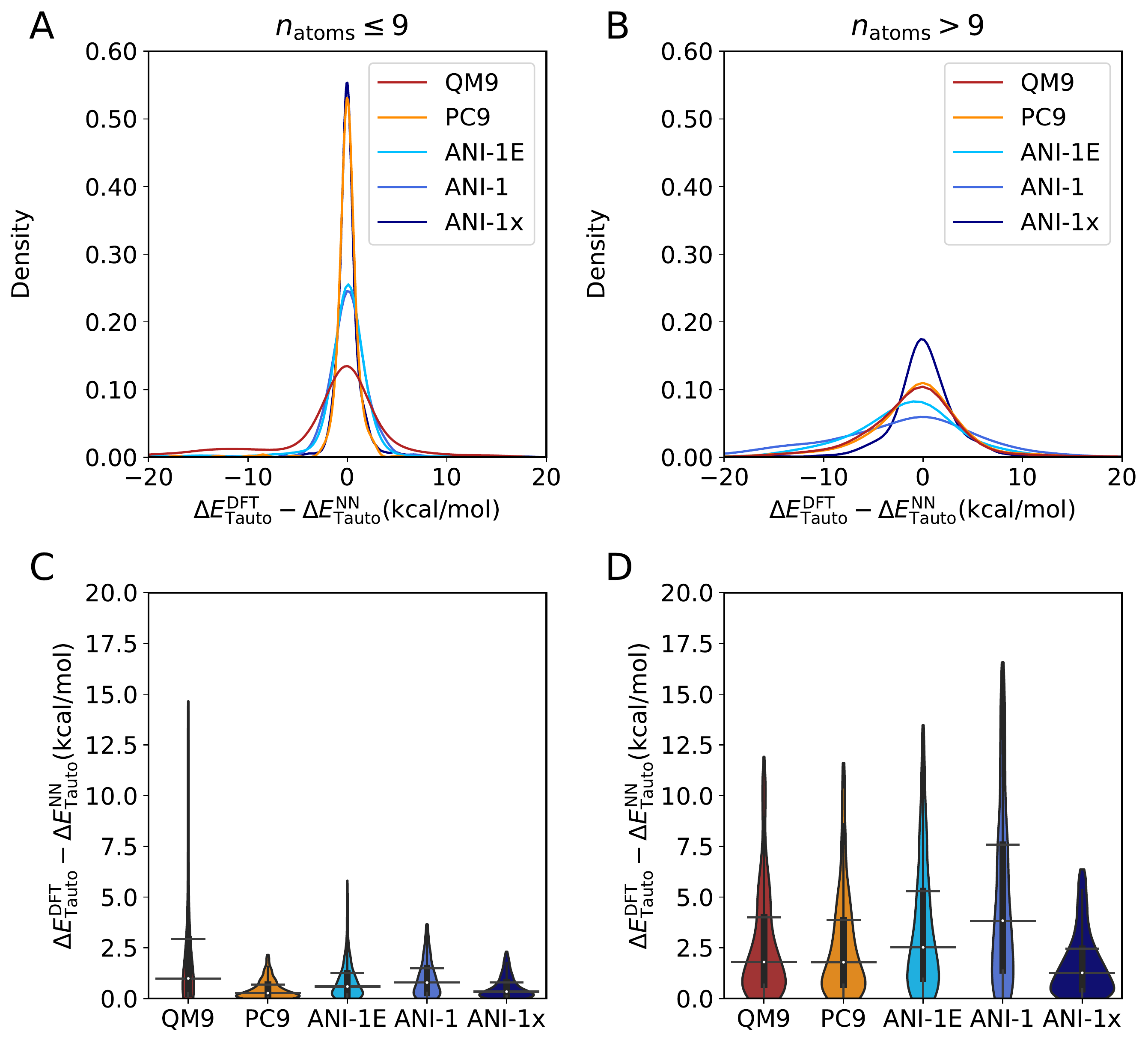}
    \caption{Error analysis on the prediction of the tautomerization
      energies. Panels A and B: Kernel density estimate of the error
      on the prediction of the tautomerization energies for the
      different databases evaluated in this work. Panels C and D:
      Normalized error distribution up to the 95\% quantile of the
      different datasets for the errors in the tautomerization
      energy. The blackbox inside spans between the 25\% and 75\%
      quantiles with a white dot indicating the mean of the
      distribution. The whiskers indicate the 5\% and 95 \%
      quantiles. The panels on the left and right are for Set1 and
      Set2, respectively.}
    \label{fig:error_dist}
\end{figure}

Next, prediction errors for $\Delta E_{\rm Tauto}$ and $E_{\rm SI}$
are analyzed and discussed for all trained models. In
Figures~\ref{fig:error_dist}A and B, the kernel density estimate of
the error distribution is reported. The violin plots in
Figures~\ref{fig:error_dist}C and D show the spread of errors, which
helps to identify large outliers.  \\

\noindent
In all cases studied, the distribution of errors for $\Delta\Delta
E_{\rm Tauto} = \Delta E_{\rm Tauto}^{\rm DFT} - \Delta E_{\rm
  Tauto}^{\rm NN}$ is centered around zero. The width of the
distribution depends on the reference dataset and on the number of
atoms in the molecule. For Set1 (Figure~\ref{fig:error_dist}A), the
error distributions for PC9 and ANI-1x suggest high probabilities
($p(\Delta \Delta E_{\rm Tauto})= p(\Delta E_{\rm Tauto}^{\rm DFT} -
\Delta E_{\rm Tauto}^{\rm NN})$), around 60 \%, to obtain a small
error. Conversely, QM9 performs worst with a maximum height of only
$p(\Delta \Delta E_{\rm Tauto}) \sim 15$ \% and a faint maximum below
$\Delta \Delta E_{\rm Tauto} = -10$ kcal/mol to predict such energy
differences with a larger probability than a positive value. The error
distributions for ANI-1E and ANI-1 are similar in shape which indicates
that their performance is comparable although the number of structures
in ANI-1E is one order of magnitude smaller than that in ANI-1. Hence,
adding additional structures (ANI-1 vs. ANI-1E) does not necessarily
improve performance.\\

\noindent
On the other hand, for Set2 (Figure~\ref{fig:error_dist}B), the
performance of QM9 and PC9 is comparable given the similar shape of
their distribution of $p(\Delta \Delta E_{\rm Tauto})$, ANI-1x gives
the best predictions with a maximum height of around 15 \% to obtain
an error for $\Delta E_{\rm Tauto}$ close to zero. All other reference
data sets perform inferior with ANI-1 reaching only a 5 \% $p(\Delta
\Delta E_{\rm Tauto})$ for a prediction close to zero. In addition,
for most of the data sets the error distribution is asymmetric with an
increased probability to predict a negative value for $\Delta \Delta
E_{\rm Tauto}$ compared to a positive value.  \\

\noindent
Results for the normalized error distributions $p(\Delta\Delta E_{\rm
  Tauto})$ are shown in Figures~\ref{fig:error_dist}C and D. For Set1,
PC9 and ANI-1x show the smallest outliers by magnitude with an error
below 2.5 kcal/mol. On the other hand, QM9 has the largest outliers
with some errors larger than 15 kcal/mol. The average error for all
reference distributions is around or below 1 kcal/mol for the 75 \%
quantile. For molecules in Set2, ANI-1 has the largest outliers,
followed by ANI-1E, QM9, PC9, and ANI-1x performing best with a maximum
error of around 5 kcal/mol, see Figure \ref{fig:error_dist}D.\\

\noindent
For completeness, error distributions $p(\Delta E_{\rm SI}) = p(E_{\rm
  SI}^{\rm DFT} - E_{\rm SI}^{\rm NN})$ for individual molecules and
their normalized variants are also reported in
Figures~\ref{sifig:error_dist}A to D. For Set1, the distributions for
PC9 and ANI-1x are centered around zero with peak heights at 80 \%
which decreases to 25 \% for ANI-1E. For ANI-1 it is shifted to
negative and for QM9 to positive values. For Set2
(Figure~\ref{sifig:error_dist}B), all error distributions are
asymmetric and extend to large negative values of $\Delta E_{\rm
  SI}$. The best and worst performing reference distributions are
ANI-1x and ANI-1, respectively. The normalized error distributions
(Figures \ref{sifig:error_dist}C and D) for both sets are strongly
peaked. For Set1 (Figure \ref{sifig:error_dist}C) the maxima for PC9
(2.5 kcal/mol) is the lowest whereas ANI-1 has the largest errors. For
Set2 (Figure \ref{sifig:error_dist}D), the outliers are even more
pronounced with $|\Delta E_{\rm SI}|>100$ kcal/mol for ANI-1E and
ANI-1. In general, the performance of ANI-1E is better than that of
ANI-1 with a smaller MAE, outliers of smaller magnitude and a more
compact distribution. These results are surprising, given the large
difference between the size of both datasets (ANI-1E ($\approx 57$k)
and ANI-1 ($\approx 20$M)) and confirm the earlier observation that
addition of new structures to a database does not necessarily improve
performance.\\

\noindent
In summary, for Set1 the database with broader chemical diversity
(PC9) and the database with the widest sampling of chemical and
conformational space (ANI-1x) perform best. Hence, chemical diversity
is essential for faithful prediction of $\Delta E_{\rm Tauto}$ but it
can be substituted to some extent with adequate sampling of
conformational space. For larger molecules (Set2), the best results
are obtained by ANI-1x which suggests that sampling of conformational
space improves extrapolation to larger molecules. Datasets containing
only equilibrium structures perform similarly for predicting $\Delta
E_{\rm Tauto}$.\\

\section{Effect of different database characteristics on predictions}
\label{sec:effects}
This section analyzes the predictive power of the NNs trained on the
five different training databases for tautomerization energies by
considering various chemical properties such as the number of heavy
atoms, the number of atoms of a given element, or the type of chemical
bonds. Given the non-linear nature of the NN, the relationships
between these characteristics and whether/how they are related to the
performance of the model is a challenging task. The features studied
here were selected because they might be considered for the selection
of a training database for the prediction of a chemical property (in
this case the tautomerization energy) or because they can be optimized
for the generation/enhancement of datasets used to train models for
specific purposes.\\
  
\subsection{Number of atoms}
\begin{figure}[h!]
    \centering
    \includegraphics[scale=0.45]{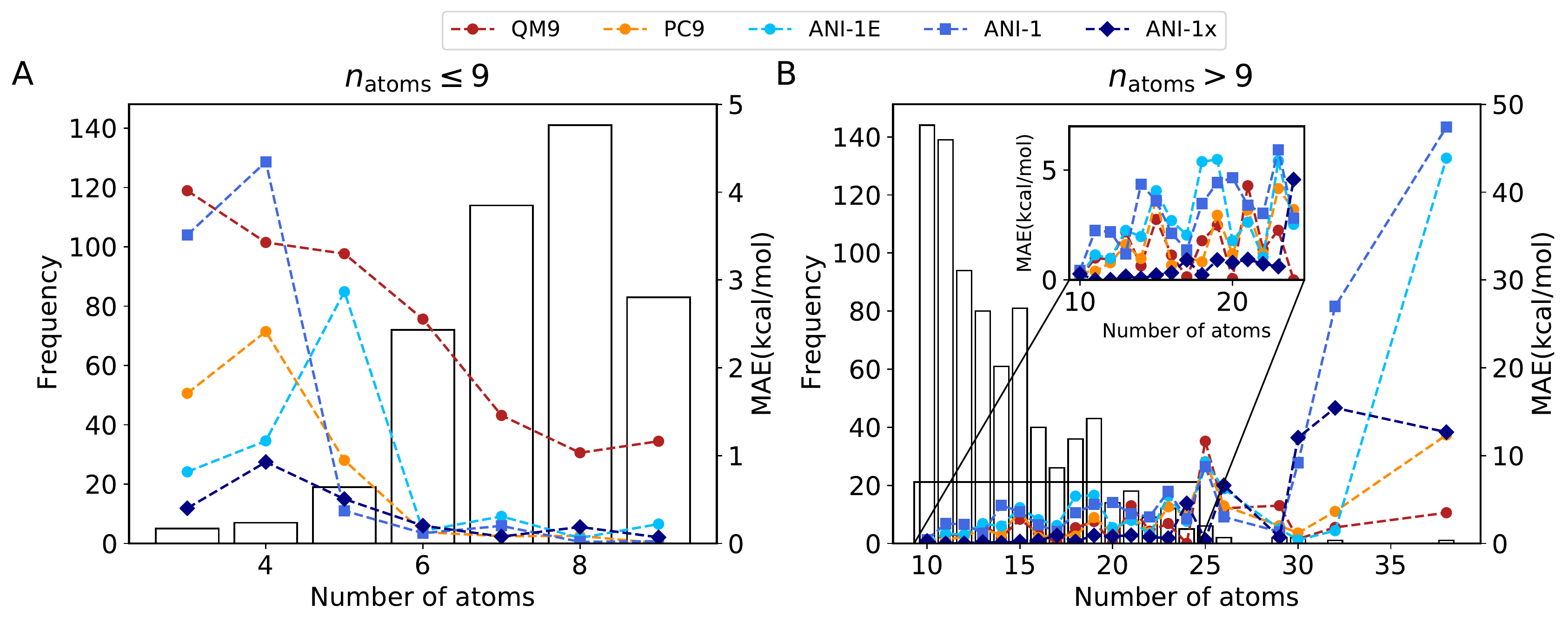}
    \caption{Mean Absolute Error (MAE) by number of heavy atoms
      (C,O,N) on the molecule for the tautomerization energy. The
      number of molecules for increasing number of heavy atoms is
      shown as a histogram. Panel A: Results for Set1, i.e. molecules
      with $n_{\rm{atoms}} \leq 9$. Panel B: for Set2. The inset in
      panel B shows the MAE for $\Delta E_{\rm Tauto}$ for $10 \leq
      n_{\rm atoms} \leq 25$.}
    \label{fig:MAEvNAtom_opt}
\end{figure}

The first characteristic considered was the number of heavy atoms (C,
N and O) contained in the reference data sets and how this affects the
prediction quality on Tautobase. For Set1 the MAE for $\Delta E_{\rm
  Tauto}$ typically decreases with increasing molecular size, see
Figure \ref{fig:MAEvNAtom_opt}A, for all five reference data sets. This
can be broadly related to the increase in the number of molecules with
the number of heavy atoms contained in the reference databases used for
training (see Figure~\ref{sifig:NatomsperDatabases}). For all data
sets except for QM9, the MAE decreases to levels below 0.5 kcal/mol as
the number of heavy atoms increases.\\

\noindent
For larger molecules (Set2) in the Tautobase, the MAEs increase
significantly, see Figure \ref{fig:MAEvNAtom_opt}B. Broadly speaking,
for up to 25 heavy atoms the MAE is still within 5 kcal/mol but
increases considerably for larger molecules. ANI-1x performs best with
${\rm MAE} < 1$ kcal/mol up to $n_{\rm atoms} = 25$ but errors
increase above 10 kcal/mol beyond that. This is followed by QM9 and
PC9 which, on average, have MAEs of $\sim 2$ kcal/mol followed by
ANI-1E and ANI-1.\\

\noindent
The MAE for predicting $E_{\rm SI}$ for Set1
(Figure~\ref{sifig:MAEvNAtom_mol}A) shows a slightly different profile
than for $\Delta E_{\rm Tauto}$ in that databases containing only
equilibrium structures show large errors for molecules with 3 and 4
heavy atoms. With increasing size the error decreases. This is most
pronounced for PC9 which eventually achieves the same quality as
ANI-1x. It should be noted that for $E_{\rm SI}$ the overall shape of
the profile of MAE vs. $n_{\rm atoms}$ for databases which only
contain equilibrium structures is similar but the magnitude of the MAE
differs. This is a consequence of the chemical diversity of the
databases as discussed in subsection \ref{subsec:vis}. For ANI-1 and
Set1, the MAE is smallest for $n_{\rm atoms} = 5$ and then starts to
grow again. For Set2 (Figure~\ref{sifig:MAEvNAtom_mol}B), the MAE
displays a steady increase with the number of heavy atoms.\\

\noindent
In summary, ANI-1x performs best across all values of $n_{\rm atoms}$
for the Tautobase, followed by PC9 across most values for $n_{\rm
  atoms}$. For Set2, QM9 is quite reliable whereas ANI-1 and ANI-1E
perform worst which reiterates the earlier finding that adding
perturbed structures to a data set does not necessarily improve the
quality on the task at hand (which is the estimation of $\Delta E_{\rm
  Tauto}$). Consequently, the results can be worse than those obtained
when training only on equilibrium structures. \\

\begin{figure}[h!]
    \centering
    \includegraphics[scale=0.5]{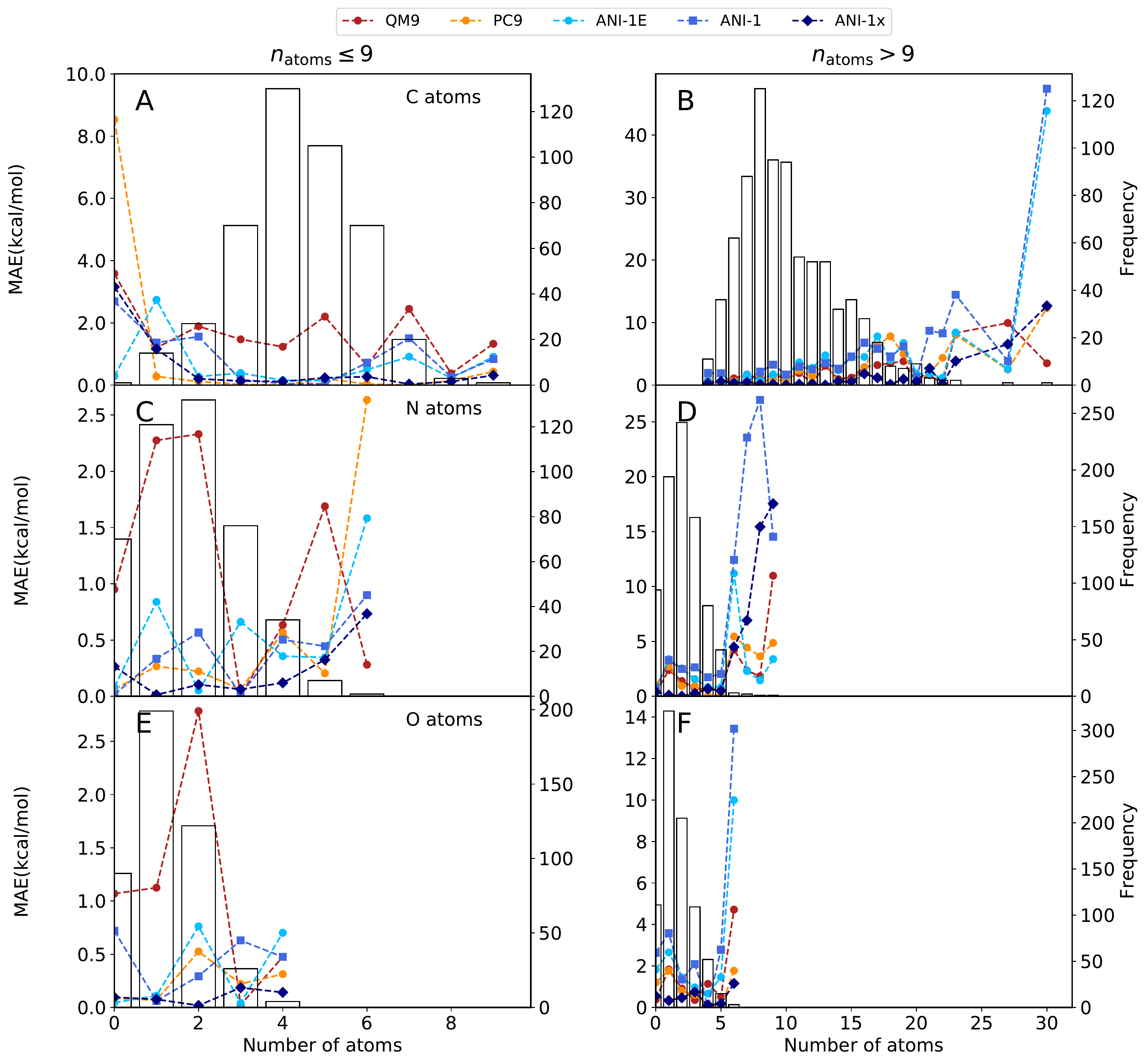}
    \caption{Mean Absolute Error (MAE) by number of atoms of a given
      element for the tautomerization energy. A histogram of the
      number of molecules for different numbers of heavy atoms is
      shown in the background. Panels A and B show the results by
      number of carbon atoms. Panels C and D shows the results by
      number of nitrogen atoms. Finally, panels E and F show the
      results by number of oxygen atoms. Left panel shows results for
      molecules with $n_{\rm{atoms}} \leq 9$ the right for molecules
      with $n_{\rm{atoms}} > 9$.}
    \label{fig:natomselementvsmae}
\end{figure}

\noindent
A further refinement can be made by analyzing the predictions for
$\Delta E_{\rm Tauto}$ in terms of the number of C-, N-, and O-atoms
contained in the molecules of the reference database (Tautobase), see
Figure \ref{fig:natomselementvsmae}. For Set1 the prediction error
tends to decrease (except for QM9) with increasing number of carbon
atoms as shown in Figure \ref{fig:natomselementvsmae}A whereas for
Set2 it increases to different extents depending on the reference
database considered, see Figure \ref{fig:natomselementvsmae}B. For
nitrogen and oxygen atoms and ANI-1x all MAEs for Set1 and Set2 are
small ($\sim 1$ kcal/mol), except for the largest numbers of N-atoms,
see Figure \ref{fig:natomselementvsmae}F. For the PC9, ANI-1E, and
ANI-1 databases and Set1 all MAEs are below or around 1 kcal/mol
whereas for QM9 they can be larger. For Set2, the MAEs are up to 5
kcal/mol for molecules for which at least tens of representatives are
contained in Tautobase, but start to increase significantly below
that, see Figure \ref{fig:natomselementvsmae}D and F.\\

\noindent
Considering the MAE for $E_{\rm SI}$ confirms these general findings,
see Figure ~\ref{sifig:natomselementvsmae}. For C-atoms the MAE for
all reference databases decreases for Set1 except for $n_{\rm atoms} =
8$ for ANI-1E and ANI-1, and increases moderately for Set2 with
increasing value of $n_{\rm atoms}$ for QM9 and ANI-1x and more
steeply for PC9, ANI-1E and ANI-1 (see Figures
~\ref{sifig:natomselementvsmae}A and B). For nitrogen and oxygen atoms
satisfactory performance is only found for PC9, ANI-1x and ANI-1E
(Set1) and for QM9 and ANI-1x (Set2), see Figures
~\ref{sifig:natomselementvsmae}C to F. Molecules with a small number
of atoms of a given element have a reduced number of different
chemical environments (see Figure
~\ref{sifig:NatomselementperDatabases}). This makes it more difficult
to predict $\Delta E_{\rm Tauto}$ if that chemical environment is
present in the target data set (Tautobase) but not sampled in the
reference sets. Consequently, larger errors are observed for molecules
with few atoms of a given element.\\

\subsection{Structural composition of the chemical databases}
The structural diversity of the databases can also be quantified in
terms of the bond types that are covered. It can be assumed that the
NN model learns that specific composition, and consequently, if the
database used for training a NN model covers a large range of bond
lengths, better results are expected. In the following, bond length
distributions in the reference databases of equilibrium structures
(PC9, QM9, and ANI-1E) are compared with the distributions contained in
Tautobase. Figures~\ref{sifig:dist_bonds_C_9} to
\ref{sifig:dist_bonds_N_9p} show that the reference and target
distributions have a different coverage of bond lengths. The general
finding is that for Set1 the overlap between reference and target
distributions is better than for Set2.\\

\noindent
Figure~\ref{sifig:dist_bonds_C_9} shows that C-C single bonds between
C(sp$^3$) atoms are well covered for the three reference databases
compared with Tautobase. The C(sp$^2$)-C(sp$^2$) double bonds are
covered differently for the reference datasets: QM9 has the fewest
examples of this type of bond, whereas ANI-1E shows the best
coverage. Such bonds are important for large molecules
($n_{\rm{atoms}} > 9$) because of the presence of aromatic rings
(Figure~\ref{sifig:dist_bonds_C_9p}).  Double C(sp$^2$)-C(sp$^2$)
bonds close to hetero atoms are poorly covered by all reference
datasets. Those bonds are crucial because they are the main origin of
tautomerization rearrangement.\\

\noindent
C(sp$^2$)-N double bonds (Figure~\ref{sifig:dist_bonds_C_9}) are
abundantly present in the Tautobase. However, the coverage of the
reference datasets of that type of bond is heterogeneous; ANI-1E shows
the best coverage followed by PC9 and QM9. On the other hand,
C(sp$^2$)-N bonds close to a heteroatom, more prevalent in larger
molecules, are better covered by QM9 than PC9 whereas C(sp)-N bonds
are well covered by all three databases. Carbon-Oxygen bonds for
carbonyl groups are more predominant in Set1 and are well covered for
the reference databases. Bonds for enols, esters and others are
important for the Tautobase; PC9 covers such C-O bonds sufficiently
but it is poorly sampled for QM9. Lastly, while C-O bonds of the type
of alcohols and dialkyl ethers are most sampled for the reference
databases they are least important for the Tautobase. \\

\begin{figure}
    \centering
    \includegraphics[width=0.85\textwidth]{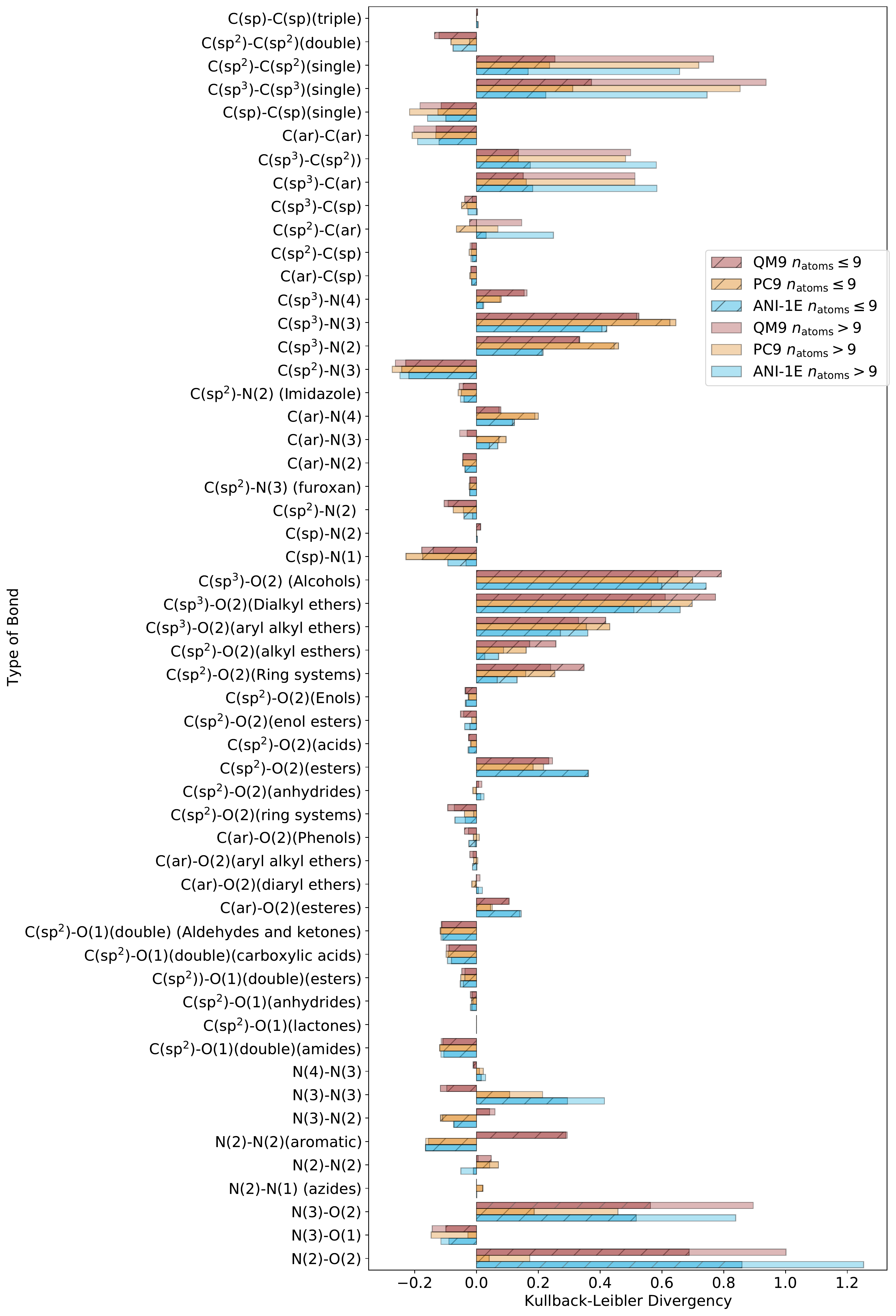}
    \caption{Values of the Kullback-Leibler divergence for different
      types of bonds present in the reference databases (QM9,PC9 and
      ANI-1E) compared with tautobase.}
    \label{fig:KL_all}
\end{figure}

\noindent
A quantitative measure for the overlap of two distributions is the KL
divergence $D(p||q)$, see Equation \ref{eq:kl}. If the two
distributions are identical, $D(p||q)=0$. On the other hand, if the
reference database $p(x)$ (here QM9, PC9, ANI-1E) contains more
information than the target set $q(x)$ (Tautobase), $D(p||q)>0$, and
if specific information is missing, $D(p||q)<0$. Hence, the cases for
which $D(p||q)<0$ are of particular relevance if improvements of the
reference databases are sought for better capturing $\Delta E_{\rm
  Tauto}$.\\

\noindent
The KL divergence analysis indicates that the coverage of the
reference sets is heterogeneous, see Tables \ref{sitab:KL_9},
\ref{sitab:KL_9p} and Figure \ref{fig:KL_all}. There are several types
of C-C bonds that are insufficiently covered, such as
C(sp$^2$)-C(sp$^2$) single and double bonds, or C(ar)-C(ar)
bonds. Also, certain types of C-N bonds would require more data as the
bonds involving C(sp$^3$) and C(sp$^2$) with different types of
nitrogens. Coverage of C-O bonds by the reference databases displays a
bias toward alcohols, ethers and esters. Finally, N-N are the types of
bond that show a more diverse coverage between databases, with some
cases for which QM9 has a good coverage (N(3)-N(2) and
N(2)-N(2)(aromatic)) but a poor coverage for N(3)-N(3). Interestingly,
there are cases for which QM9 has a good coverage, whereas ANI-1E and
PC9 are deficient. Figure \ref{fig:KL_all} shows that none of the
reference databases covers all of the predominant types of bonds
present in the tautobase.\\

\begin{figure}[h!]
    \centering
    \includegraphics[scale=0.5]{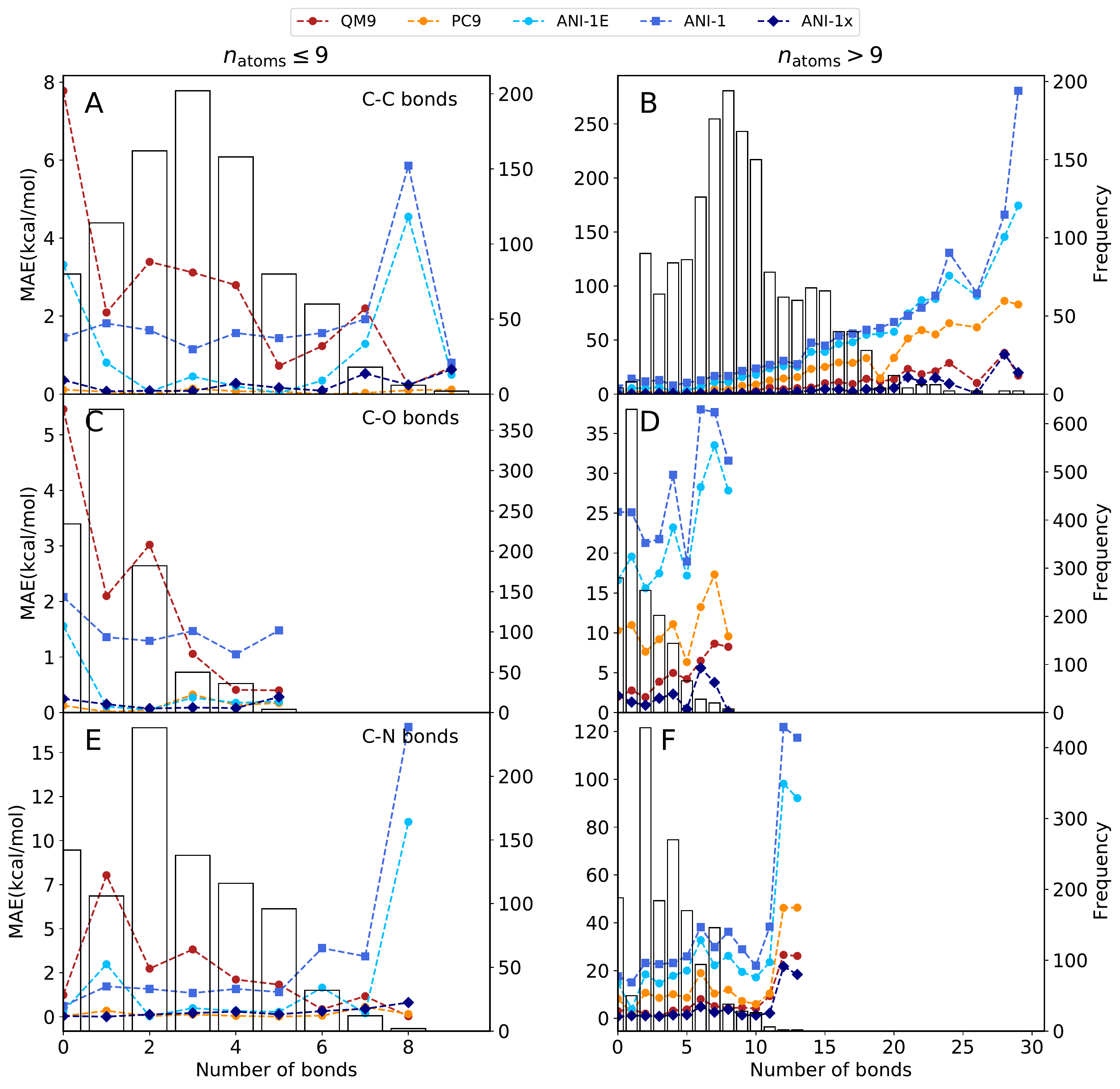}
    \caption{Mean Absolute Error (MAE) of the prediction of the energy
      of single isomers by number of bonds involving carbon atoms. The
      number of molecules for different numbers of bonds is shown as a
      histogram. Panels A and B correspond to the results of C-C
      bonds, panels C and D show the results for C-O bonds and panels
      E and F show the results for C-N bonds. Left panel for Set1 and
      right panel for Set2.}
    \label{fig:nbondsvsmae}
\end{figure}

\noindent
Next, the MAE for a specific number of a particular type of bond
(e.g. C-C, C-O, or C-N) was determined for single isomer energies, see
Figure \ref{fig:nbondsvsmae}. The results in Figure
\ref{fig:nbondsvsmae}A show that for C-C bonds and Set1 the error for
PC9 (orange) and ANI-1x (black) is constant and well below 1
kcal/mol. On the other hand, for QM9 (red) the error oscillates
without following a clear trend. ANI-1E (light blue) and ANI-1 (dark
blue) behave similarly to one another with a smaller MAE for ANI-1E
than the one for ANI-1. \\

\noindent
For C-O bonds, the MAE of the prediction of $E_{\rm SI}$ slowly
increases for PC9 but remains well below 1 kcal/mol, whereas for QM9
it starts at above 5 kcal/mol and decreases to below 1 kcal/mol but
always remaining above that for PC9, see Figure
\ref{fig:nbondsvsmae}C. The error for the database of the ANI family
is largely constant over the number of bonds. For ANI-1 the MAE
oscillates between 1 kcal/mol and 2 kcal/mol, whereas for ANI-1E and
ANI-1x the MAE is well below 1 kcal/mol, except for zero C-O bonds and
ANI-1E. Considering C-N bonds (Figure \ref{fig:nbondsvsmae}E) it is
found that their maximum number is larger than that for C-O bonds. The
magnitude for the MAE for this bond type is at least a factor of three
larger than that for the C-C and C-O bonds, respectively. Again, PC9
and ANI-1x perform best, followed by ANI-1E (except for molecules with
only one C-N bond). The MAE for QM9 slowly decreases whereas that for
ANI-1 is constant at below 2 kcal/mol up to five C-N bonds after which
it sharply increases.\\

\noindent
Regarding C-O and C-N bonds, it is clear that the good coverage of
ANI-1E helps to reach small MAE when the number of bonds increases
(Figure \ref{sifig:dist_bonds_C_9}). These results show that PC9 has a
good overall performance because there is an adequate coverage of
different chemical bond types whereas QM9 and ANI-1E have biases toward
some types of bonds (Table \ref{sitab:KL_9} and \ref{sitab:KL_9p}). It should be stressed
though that such an analysis excludes the fact that the same type of
bond can behave differently given different chemical environments. \\

\noindent
For Set2 the increase of the error with decreasing number of samples
is more apparent. As discussed before, the MAE observed for larger
molecules ($n_{\rm{atoms}}>9$) grows proportional with the number of
bonds (See Figure \ref{fig:nbondsvsmae} B, D and F). In this regard,
ANI-1 and ANI-1E are the databases with largest growth rate, followed
by PC9 and finally QM9 and ANI-1x. The low MAE for $E_{\rm SI}$ by
number of bonds obtained with ANI-1x is a consequence of the addition
of an adequate number of non-equilibrium structures. This suggests
that a lack of chemical diversity can be partially compensated by
including non-equilibrium structures in a database. \\

\subsection{Initial Geometry}
In the previous sections, the energy of the molecules was computed
using the equilibrium geometry of the tautomeric pairs computed at the
level of theory of the various reference databases used for training
the NNs. However, in practice, it would be of interest to sidestep the
computationally rather expensive optimization of the structures in the
reference dataset (here Tautobase) at the density functional or even
higher level of quantum chemical theory. For this, using empirical
force fields is a possibility. A recent study for 3271 small organic
molecules ($n_{\rm atoms}<50$), similar to those contained in
Tautobase, found typical RMSDs of 0.25 \AA\/ to 1 \AA\/ between
optimized structures at the B3LYP/6-31G* level of theory compared with
those optimized by nine different force fields, including GAFF,
MMFF94, OPLS and others.\cite{lim2020benchmark} Considering this, it
is interesting to assess the performance of the NN-based models on
FF-optimized geometries. \\

\noindent
For this analysis, the geometries for molecules from Tautobase were
optimized as described in section \ref{subsec:initial_geom} with the
MMFF94 force field and then used to evaluate the tautomerization
energies using the five trained NNs. Table~\ref{sitab:mae_MMFF94}
shows that the MAE for the tautomerization and single isomer energy
increases for all evaluated models when the geometry used to evaluate
the energies differs from geometries optimized with the respective
{\it ab initio} method. In all cases the MAE for $\Delta E_{\rm
  Tauto}$ increases by a factor between 1.5 and 3 compared with the
error obtained using the optimized geometries at the quantum chemical
level of theory used to train the NN (see Table \ref{tab:mae}). A
similar effect is observed for $E_{\rm SI}$. It is noticeable that
this geometry effect is less pronounced for databases which contain
non-equilibrium structures: ANI-1 shows the smallest increase of the
MAE for $\Delta E_{\rm Tauto}$ compared with results from using
optimized geometries at the appropriate level of theory.\\

\noindent
Normalized distributions for $\Delta E_{\rm Tauto}$ using MMFF94
geometries for Set1 (Figure \ref{sifig:error_dist_MMFF94}C) indicate
that the datasets which only cover chemical space (QM9, PC9 and ANI-1E)
perform similarly with the highest values of the outliers close to 15
kcal/mol. Conversely, the other two databases (ANI-1 and ANI-1x) have
a more compact distribution with the maximum values for outliers
around 10 kcal/mol. Figure~\ref{sifig:error_dist_MMFF94}D shows that
the most challenging case for predicting tautomerization energies is
for Set2 with geometries generated with MMFF94. However, the
performance is similar for all the datasets evaluated with outliers
larger than 20 kcal/mol, except for ANI-1x with a maximum of 10
kcal/mol. These results demonstrate that the initial geometry passed
to the NN model is essential for obtaining meaningful results. Scoring
a model trained on minimum energy structures computed at a given level
of quantum chemical theory can not be done using optimized structures
from an empirical force field (or from structures at a sufficiently
different level of quantum chemical theory).\\

\noindent
To confirm this finding, a third set of molecules was evaluated using
geometries generated by six popular force fields as described in
Section \ref{sec:methods}. In the SAMPL2
challenge,\cite{geballe2010sampl2} the RMSD to DFT optimized
geometries of several molecules were evaluated with respect to
geometries obtained from various force fields. The energy predictions
for the molecules on the SAMPL2 challenge (Figure
\ref{sifig:RMSD_scatter}) show no correlation between the geometry and
the energy predicted by the NN. There are molecules with a small RMSD
(e.g. $\leq 0.1$ \AA) which display a significant error ($>5$
kcal/mol) in predicting the energy by the NN models and vice versa. A
possible explanation is that the change in geometry can be compensated
by more extensive sampling of chemical space by the reference
databases.\\

\subsection{Visualization of chemical space}
\label{subsec:vis}
\begin{figure}[h!]
    \centering
    \includegraphics[scale=0.45]{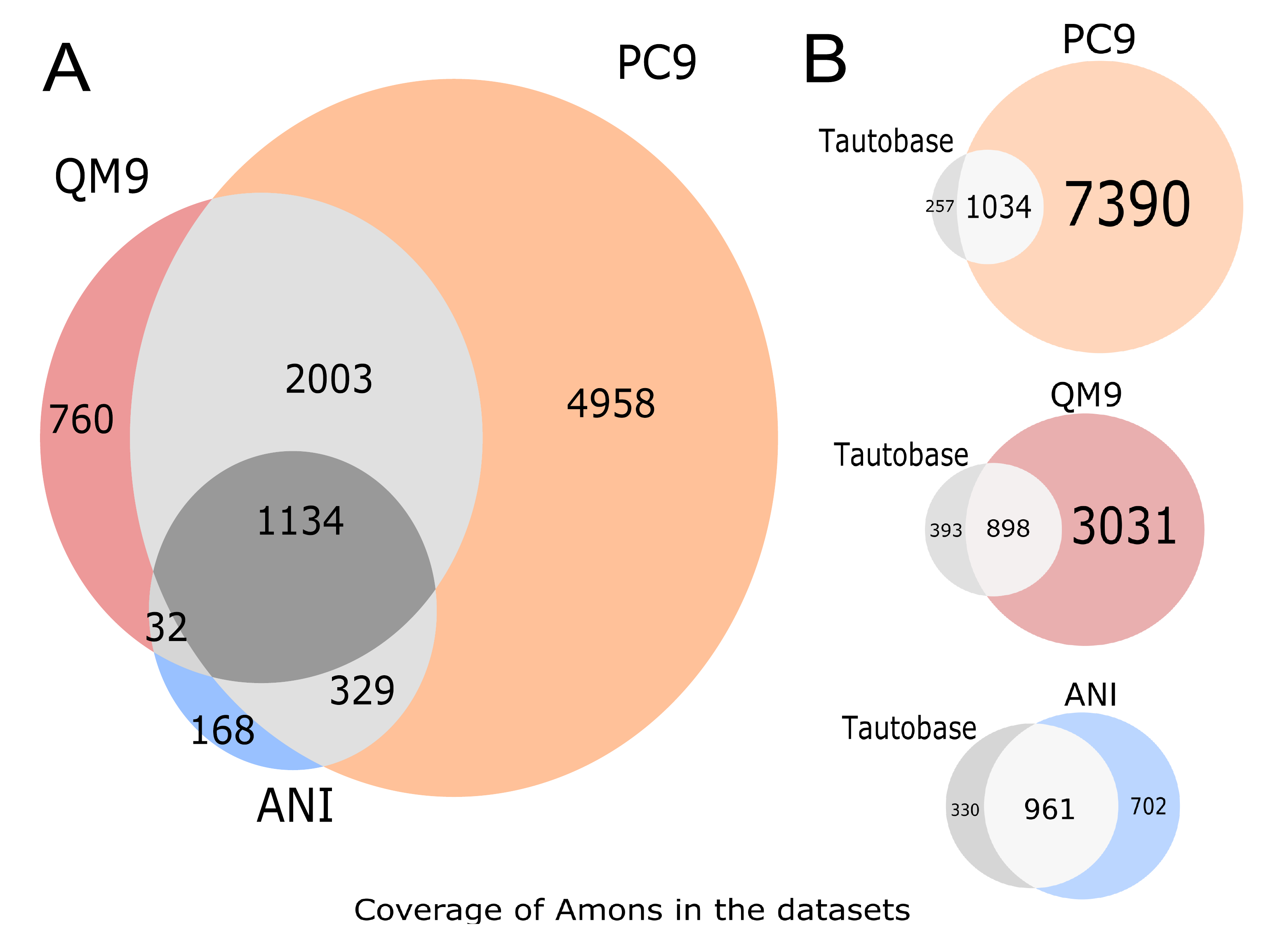}
    \caption{A) Venn diagram showing the overlap of amons between the
      QM9\cite{ramakrishnan2014quantum},
      PC9\cite{glavatskikh2019dataset} and ANI\cite{smith2017ani}
      family data sets. B) Overlap of amons up to length five between
      three popular quantum chemistry datasets (QM9, PC9 and ANI-1E)
      and the `Tautobase'\cite{wahl2020tautobase}, a collection of
      experimentally observed tautomers. Molecules containing other
      atoms than hydrogen, carbon, nitrogen, and oxygen were filtered
      from all datasets. }
    \label{sifig:venn}
\end{figure}

To understand the influence of the different databases studied on the
performance of the models, it is of interest to analyze the coverage
of `chemical space'.  Firstly, molecules in the databases were
deconstructed and their constituent amons (unique chemical fragments)
were enumerated.  PC9 contained the largest number of unique amons
(8424), followed by QM9 (3929) and, finally, the ANI family of
databases (1663). There is significant overlap of common amons between
the datasets (Figure \ref{sifig:venn}A, which suggests that they cover
similar regions of chemical space. Regarding the overlap of the test
set (Tautobase) with the databases tested, PC9 is the one which covers
the most amons by number in the reference set, followed by ANI-1E and,
QM9 (Figure \ref{sifig:venn}B).\\

\noindent
It is of interest to quantify the overlap of amons between the
reference and target sets as done in figure \ref{sifig:venn}. This
provides a measure for the coverage of chemical space common to both
sets and can be related to prediction errors for $\Delta E_{\rm
  Tauto}$. Comparing this overlap with the results obtained in
\ref{tab:mae}, the database that covers the largest number of amons
gives the smallest MAE for $\Delta E_{\rm Tauto}$. This is more
evident if the MAEs for $\Delta E_{\rm Tauto}$ for Set1 and Set2 are
analyzed separately. For Set1, the results are consistent with the
classification by number of amons (Figure \ref{fig:scatter_chemspace}
A, C and E). However, for Set2 the results of QM9 are better than
those obtained with ANI-1E which contradicts the correlation with the
number of amons in the databases (Figure \ref{fig:scatter_chemspace}
B, D and F). This can probably be explained by the fact that QM9
contains more large molecules which should help predicting the energy
of single molecules accurately. This effect is also observed looking
at the MAE for $E_{\rm SI}$ in Table \ref{tab:mae} for which QM9 has a
smaller MAE than PC9.\\

\begin{figure}[h!]
    \centering
    \includegraphics[width=0.9\textwidth]{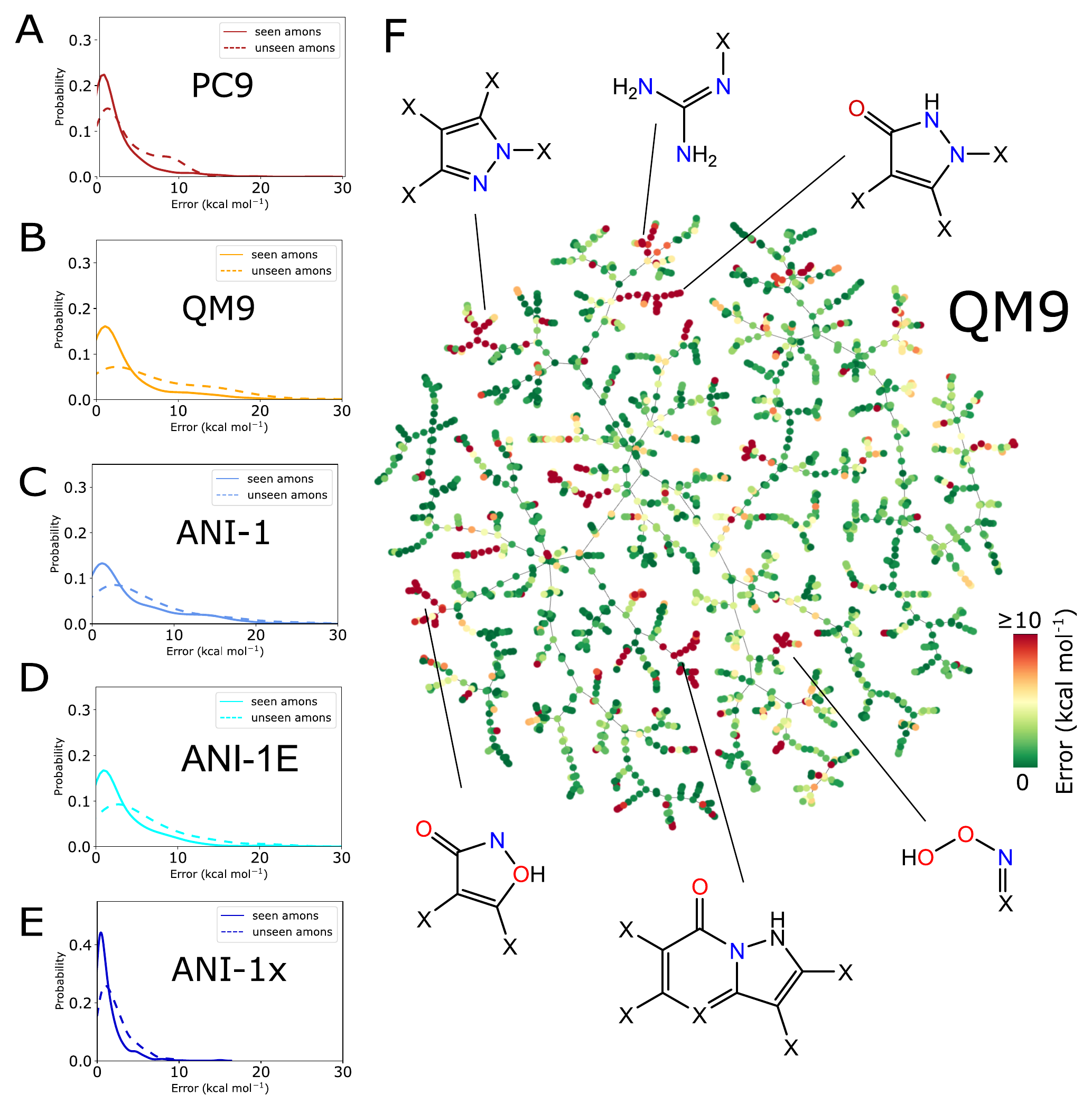}
    \caption{The relationship between chemical space and model
      reliability. Panels A to E: The mean average absolute error for
      molecules from Tautobase with one or more amons outside the
      training set is larger than if all amons are present, a trend
      observed for all databases. Panel F: Projection of the Tautobase
      using the TMAP algorithm identifies `branches' of chemical space
      that are poorly predicted by the neural network models. The
      error displayed here is for the QM9 database. TMAPs for all
      databases used for training are available as interactive plots
      which can be viewed in a web browser, which can be obtained in
      the supporting information.}
    \label{fig:tmap}
\end{figure}

\noindent
A more detailed analysis is possible by considering if the amons of
the isomers in the tautobase are present (or not) in the training
databases. For this, a set `seen amons' (all constituent amons
included in the reference database) and `unseen amons' (molecules for
which one or several amons were missing from the reference set) was
defined. The error distributions for both sets were determined and are
reported in Figure \ref{fig:tmap}. Perhaps unsurprisingly, the `seen
amons' had a larger probability of obtaining a small error compared to
the `unseen amons'.  Interestingly, PC9, which provides the broadest
sampling of chemical space as quantified by the number of amons in the
database, showed a similar probability error distribution for Set1 and
Set2. The errors for `unseen amons' using the NN trained on QM9, a
significantly smaller dataset, shows a larger and more right-skewed
distribution of errors. One possible explanation may be that a better
exploration of chemical space helps when predicting energies for
molecules containing chemistry outside that covered by the database.\\

\noindent
Regarding the ANI family of databases, the ANI-1 and ANI-1E results
have similar error distributions. However, ANI-1x shows a smaller mean
error for both the seen and unseen sets.  These results are another
indication that a random sampling of conformational space does not
help improve the NN model predictions. On the contrary, it makes it
worse than when only equilibrium structures are considered. Another
notable finding is that ANI-1x shows similar performance for molecules
with seen and unseen amons. This can be explained given the good
sampling of chemical space, which is the same as for ANI-1E, but
combined with a broad exploration of conformational space by a
refinement from ANI-1 using active learning \cite{smith2018less}.\\

\noindent
Rational detection of systematic deficiencies in quantum chemical
databases is challenging because of the high dimensionality of
chemical space. For this reason, methods to visualise chemical space
in a digestible way are highly desirable. A recent development is the
TreeMAP (TMAP) algorithm, which allows for an interpretable, low
dimensional representation of the test set's chemical
space\cite{probst2020visualization}. The TMAP algorithm constructs a
weighted graph that is efficient for compact representations of high
dimensional data. The necessary weights are based on the Jaccard
distance, which measures the dissimilarity between the fingerprint of
two structures. This graph is then pruned to the minimum spanning
tree, a fully connected, acyclic sub-graph containing all nodes of the
parent graph, and retaining only essential edges which minimize the
weights. This organizes the compounds contained in a database into a
tree, putting related structures on nearby branches. From this, groups
of related moieties can be identified which are potentially
detrimental to predicting the quantity in question, here $\Delta
E_{\rm Tauto}$.\\

\noindent
For instance, coloring the nodes of Tautobase TMAP by error of
tautomerization energy (Figure \ref{fig:tmap}F) reveals that
structures with azoles containing N-N and N-O bonds correlate with
large errors. Interestingly, KL-divergences for these types of bond
distances suggested that they were underrepresented in the reference
sets, see Figure \ref{fig:KL_all}. The moieties corresponding to large
errors change based on the different databases used to train the NN
model, see Figure \ref{sifig:TMAP_SI}. Interactive plots are available
in the supporting information.\\

\section{Discussion and Conclusions}
\label{sec:conclusions}
\noindent
The prominence of ML has raised concerns regarding the
'interpretability' of the models
conceived\cite{lipton2018mythos,dybowski2020interpretable}. This
awareness also increases for complex models because a rational
relationship between initial data used for training and resulting
prediction becomes less transparent. Therefore, it is important to
develop quantifiable and intuitive tests for how ML models ``work''
and how trustworthy predictions by them
are\cite{schutt2019quantum}. One recently proposed procedure is
``post-hoc'' interpretation for which the practitioner analyzes a
trained model with the aim to understand what the model has learned
from the data without changing the underlying
model\cite{murdoch2019definitions,du2019techniques}. In this work,
post-hoc interpretability techniques were used to investigate the
effect of different features of the database on predicting a chemical
property (tautomerization energy). The selected features are
considered important for the construction of robust quantum chemical
databases for ML. In the present case this implied the analysis of
individual features of several databases to quantify how these modify
the prediction of a chemical property on an unseen set of examples
using statistics and visualization techniques. With sufficient
information from such an analysis it is expected that it will be
possible to identify which features of the training databases are
essential for good performance on a given task, making a rational
design/enhancement of databases for training ML models for a given
task possible.\\

\noindent
The present work aimed at quantifying and analyzing the suitability of
NNs trained on five different reference data sets (QM9, PC9, ANI-1E,
ANI-1, and ANI-1x) to predict the tautomerization energies of
molecules contained in Tautobase. It was found that depending on which
characteristics are considered, the predicted MAEs can behave very
differently and can, in part, be related to geometrical and/or
chemical properties encoded in the databases. Such analyses attempt to
digress from ``black box'' applications of ML methods and move towards
``interpretable ML''. Hence, one of the questions is ``what features
need to be present and covered in a training database for application
to a concrete chemical question''. In the present case the databases
to choose from were QM9, PC9, ANI-1E, ANI-1, and ANI-1x' and the
application was computation of the gas phase tautomerization energy.\\

\noindent
The results indicate that the exploration of chemical space is
essential for meaningful results. The coverage of chemical space can
be quantified by the chemical diversity expressed as the number of
amons on the database (see Figure \ref{sifig:venn}). The energy
prediction improves when the overlap of the number of amons in the
training set and the tautobase increases. If the number of amons in
the chemical database does not cover all the amons on the target set,
addition of non-equilibrium structures to the training set can improve
the results. That addition needs to be done following a rational
strategy because an arbitrary addition causes 
deficiencies in the model, leading to poor results as seen for the
results for ANI-1E and ANI-1. \\

\noindent
Another determinant property is the number of heavy atoms in molecules
covered in the database. Not surprisingly, better results are obtained
for the range covered by the database. Outside that range, the energy
prediction quality decreases with the number of atoms for most
databases. One of the training databases (ANI-1x) shows good results
because the non-equilibrium structures help in predicting the
energies. It was observed that the different chemical environments
need to be thoroughly sampled because functional diversity is key to
assure good results.\\

\noindent
The structural composition of the data sets used for training the NNs
(QM9, PC9 and ANI-1E) and the data set to which the trained models
were applied to (Tautobase) can be compared through the
Kullback-Leibler divergence. The overlap between these distributions
already provides an indication how suitable a particular reference
data set will be for application to the target task. In other words:
the KL divergence can be used for the rational design of databases for
NN models. It will be of interest to extend this to angles and
dihedrals for a comprehensive exploration of the structural overlap.
The geometry of the molecule to evaluate the trained NN is essential
for good performance. Using the TMAP algorithm it was possible to
identify regions of chemical space that are poorly covered by the
trained models. This is an excellent aid because it becomes easier to
improve sampling of a specific region of chemical space to obtain
better results. The characteristics of the databases analyzed in the
present work can be used as a rational basis to determine whether a
database is suitable for a specific prediction task.\\

\noindent
In conclusion, the present work demonstrates that ML-trained models on
five different reference databases and applied to one specific task
(tautomerization energy) perform with a MAE ranging from 1.7 kcal/mol
to 4.6 kcal/mol. The best performing reference database (ANI-1x with 5
M structures) performs on average by 1 kcal/mol better than PC9 which
contains about two orders of magnitude fewer reference structures
($\approx 85$ K). On the other hand, PC9 is chemically more diverse by
a factor of 5 (as judged from the number of amons) compared with the
ANI family of databases. This indicates that lack in chemical
diversity can be compensated for by increased number of
non-equilibrium structures. However, the scaling of these two
properties is very different. Together with quantitative descriptors,
such as the KL divergence, the present results and analyses suggest
that a rational approach to database generation for specific tasks may
be possible.\\

\section*{Data Availability Statement}
The PhysNet codes are available at Github
(\url{https://github.com/MMunibas/PhysNet}) . The database for ANI-1E
(\url{10.5281/zenodo.4680953}) and the geometries used for the
tautobase (\url{10.5281/zenodo.4680972}) are available at Zenodo.
    
\section*{Acknowledgment}
The authors acknowledge financial support from the Swiss National
Science Foundation (NCCR-MUST and Grant No. 200021-7117810) and the
University of Basel. OTU acknowledges funding from the Swiss National
Science Foundation (Grant No. P2BSP2\_188147).

\newpage

\bibliography{biblio1}

\newpage

{\bf Table of Content}

\begin{figure}[H]
  \centering \includegraphics[width=0.50\linewidth]{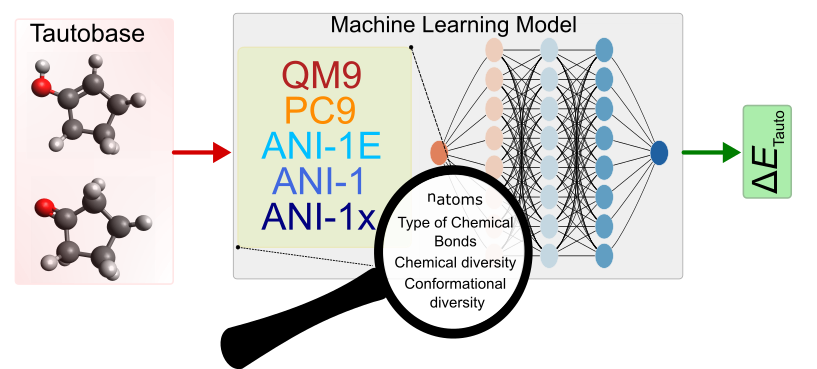}
    \label{fig:toc}
\end{figure}

\end{document}


\date{\today}

\section{Supplementary Methods}

\subsection{Generation of the geometries for the molecules of the SAMPL2 challenge}
The molecules of the SAMPL2 challenge\cite{geballe2010sampl2} were
used to generate a third database for the evaluation of the energies
by the NN models containing only equilibrium molecules. The initial
geometries generation was done using OpenBabel\cite{o2011open} for the
force fields GAFF\cite{wang2004development}, UFF\cite{rappe1992uff}
and Ghemical\cite{hassinen2001new} for 5000 steps from the SMILES
representations. \\

\noindent
For the Gromos\cite{schmid2011definition} force field, the
parametrization of the molecules was done using the ATB
server\cite{koziara2014testing}. Meanwhile, the parameters for the
CHARMM\cite{foloppe2000all} force field were generated using
SwissParam\cite{zoete2011swissparam}. Finally, the parameters of the
OPLS\cite{jorgensen1988opls} force field were generated using the
LigParGen\cite{dodda2017ligpargen} code. Once the parameters from
those force fields were generated, the molecules were optimized for
5000 steps using the GROMACS code.

\newpage

\begin{longtable}[c]{|l|c|c|c|c|}
\caption{Values of the Kullback-Leibler (KL) divergence of the
  different training sets and the tautobase at specific intervals for
  different types of bonds for molecules with $n_{\rm{atoms}} \leq
  9$. The values of the bond lengths were taken from Allen, F.H.,
  \textit{et al.}, 2006 \cite{allen2006typical} }
\label{sitab:KL_9}
\\\hline
Type of Bond                                                                                                                                         & QM9     & PC9     & ANI-1E    & Interval(\r{A})  \\ \hline
\endfirsthead
\caption{Values of the Kullback-Leibler (KL) divergence of the
  different training sets and the tautobase at specific intervals for
  different types of bonds for molecules with $n_{\rm{atoms}} \leq
  9$. The values of the bond lengths were taken from Allen, F.H.,
  \textit{et al.}, 2006 \cite{allen2006typical}(cont.) } \\\hline Type
of Bond & QM9 & PC9 & ANI-1E & Interval (\r{A}) \\ \hline \endhead
\hline \multicolumn{4}{r}{\textit{Continued on next page..}}
\\ \endfoot \hline \endlastfoot

C(sp)-C(sp)(triple)                                                                                                                                  & 0.003   & 0.000   & 0.005   & 1.167-1.197 \\ \hline
C(sp$^2$)-C(sp$^2$)(double)                                                                                                                                & -0.136  & -0.083  & -0.076  & 1.280-1.405 \\ \hline
C(sp$^2$)-C(sp$^2$)(single)                                                                                                                                & 0.254   & 0.236   & 0.167   & 1.400-1.568 \\ \hline
C(sp$^3$)-C(sp$^3$)(single)                                                                                                                                & 0.372   & 0.312   & 0.224   & 1.458-1.610 \\ \hline
C(sp)-C(sp)(single)                                                                                                                                  & -0.114  & -0.125  & -0.099  & 1.374-1.474 \\ \hline
C(ar)-C(ar)                                                                                                                                          & -0.131  & -0.132  & -0.121  & 1.350-1.440 \\ \hline
C(sp$^3$)-C(sp$^2$)                                                                                                                                        & 0.136   & 0.135   & 0.174   & 1.470-1.538 \\ \hline
C(sp$^3$)-C(ar)                                                                                                                                         & 0.151   & 0.161   & 0.182   & 1.479-1.539 \\ \hline
C(sp$^3$)-C(sp)                                                                                                                                         & -0.038  & -0.048  & -0.027  & 1.436-1.481 \\ \hline
C(sp$^2$)-C(ar)                                                                                                                                         & -0.022  & -0.065  & 0.031   & 1.441-1.512 \\ \hline
C(sp$^2$)-C(sp)                                                                                                                                         & -0.015  & -0.016  & -0.013  & 1.425-1.441 \\ \hline
C(ar)-C(sp)                                                                                                                                          & -0.017  & -0.019  & -0.015  & 1.430-1.448 \\ \hline
                                                                                                                     \multicolumn{5}{|c|}{Carbon-Nitrogen bonds}                   \\ \hline
C(sp$^3$)-N(4)                                                                                                                                          & 0.1541  & 0.0765  & 0.0205  & 1.482-1.510 \\ \hline
C(sp$^3$)-N(3)                                                                                                                                          & 0.5254  & 0.6454  & 0.4214  & 1.446-1.572 \\ \hline
C(sp$^3$)-N(2)                                                                                                                                          & 0.3340  & 0.4596  & 0.2156  & 1.461-1.506 \\ \hline
C(sp$^2$)-N(3)                                                                                                                                          & -0.2290 & -0.2421 & -0.2186 & 1.314-1.419 \\ \hline
C(sp$^2$)-N(2) (Imidazole)                                                                                                                              & -0.0432 & -0.0488 & -0.0402 & 1.369-1.384 \\ \hline
C(ar)-N(4)                                                                                                                                           & 0.0790  & 0.2000  & 0.1227  & 1.461-1.470 \\ \hline
C(ar)-N(3)                                                                                                                                           & 0.0290  & 0.2841  & 0.2339  & 1.340-1.476 \\ \hline
C(ar)-N(2)                                                                                                                                           & -0.0080 & -0.0087 & 0.0113  & 1.422-1.442 \\ \hline
C(sp$^2$)-N(3) (furoxan)                                                                                                                                & -0.0202 & -0.0213 & -0.0216 & 1.311-1.324 \\ \hline
C(sp$^2$)-N(2)                                                                                                                                          & -0.1046 & -0.0753 & -0.0409 & 1.273-1.339 \\ \hline
C(ar)-N(3)                                                                                                                                           & -0.0897 & -0.0923 & -0.0962 & 1.325-1.369 \\ \hline
C(ar)-N(2)                                                                                                                                           & -0.0807 & -0.0801 & -0.0853 & 1.300-1.348 \\ \hline
C(sp)-N(2)                                                                                                                                           & 0.0131  & 0.0001  & 0.0024  & 1.140-1.148 \\ \hline
C(sp)-N(1)                                                                                                                                           & -0.1771 & -0.2281 & -0.0925 & 1.131-1.449 \\ \hline
                                                                                             \multicolumn{5}{|c|}{Carbon-Oxygen bonds}                    \\ \hline
C(sp$^3$)-O(2) (Alcohols)                                                                                                                               & 0.7924  & 0.6993  & 0.7431  & 1.395-1.449 \\ \hline
C(sp$^3$)-O(2)(Dialkyl ethers)                                                                                                                          & 0.7733  & 0.6979  & 0.6591  & 1.405-1.458 \\ \hline
C(sp$^3$)-O(2)(aryl alkyl ethers)                                                                                                                       & 0.4180  & 0.4313  & 0.3602  & 1.417-1.438 \\ \hline
C(sp$^2$)-O(2)\footnote{Aryl alkyl ethers, alkyl   esters of carboxilic acids, alkyl esters of alpha, beta unsaturated acids,   alkyl esterets of benzoic acid}  & 0.2570  & 0.1608  & 0.0713  & 1.435-1.501 \\ \hline
C(sp$^2$)-O(2)(Ring systems)                                                                                                                            & 0.3480  & 0.2537  & 0.1312  & 1.430-1.501 \\ \hline
C(sp$^2$)-O(2)(Enols)                                                                                                                                   & -0.0370 & -0.0262 & -0.0359 & 1.324-1.342 \\ \hline
C(sp$^2$)-O(2)(enol esters)                                                                                                                             & -0.0516 & -0.0156 & -0.0380 & 1.341-1.363 \\ \hline
C(sp$^2$)-O(2)(acids)                                                                                                                                   & -0.0245 & -0.0172 & -0.0248 & 1.279-1.320 \\ \hline
C(sp$^2$)-O(2)(esters)                                                                                                                                  & 0.2340  & 0.1834  & 0.3607  & 1.328-1.420 \\ \hline
C(sp$^2$)-O(2)(anhydrides)                                                                                                                              & 0.0068  & -0.0116 & 0.0138  & 1.379-1.393 \\ \hline
C(sp$^2$)-O(2)(ring systems)                                                                                                                            & -0.0932 & -0.0388 & -0.0697 & 1.332-1.377 \\ \hline
C(ar)-O(2)(Phenols)                                                                                                                                  & -0.0390 & -0.0100 & -0.0245 & 1.353-1.373 \\ \hline
C(ar)-O(2)(aryl alkyl ethers)                                                                                                                        & -0.0215 & -0.0096 & -0.0127 & 1.363-1.377 \\ \hline
C(ar)-O(2)(diaryl ethers)                                                                                                                            & -0.0010 & -0.0151 & 0.0059  & 1.375-1.391 \\ \hline
C(ar)-O(2)(esteres)                                                                                                                                  & 0.1042  & 0.0453  & 0.1445  & 1.394-1.408 \\ \hline
C(sp$^2$)-O(1)(double) (Aldehydes and   ketones)                                                                                                         & -0.1119 & -0.1178 & -0.1090 & 1.188-1.238 \\ \hline
C(sp$^2$)-O(1)(double)\footnote{Delocalized double bonds in carboylate anions}                                                                                 & -0.0370 & -0.0513 & -0.0399 & 1.232-1.262 \\ \hline
C(sp$^2$)-O(1)(double)(carboxylic acids)                                                                                                                & -0.0887 & -0.0917 & -0.0811 & 1.203-1.241 \\ \hline
C(sp$^2$)-O(1)(double)(esters)                                                                                                                          & -0.0477 & -0.0509 & -0.0523 & 1.181-1.207 \\ \hline
C(sp$^2$)-O(1)(anhydrides)                                                                                                                              & -0.0198 & -0.0164 & -0.0199 & 1.184-1.193 \\ \hline
C(sp$^2$)-O(1)(lactones)                                                                                                                                & 0.0000  & 0.0000  & 0.0000  & 1.187-1.209 \\ \hline
C(sp$^2$)-O(1)(double)(amides)                                                                                                                          & -0.1086 & -0.1185 & -0.1058 & 1.193-1.243 \\ \hline
                                                                                                                                                 \multicolumn{5}{|c|}{Nitrogen-nitrogen bonds}                      \\ \hline
N(4)-N(3)                                                                                                                                            & -0.0109 & 0.0094  & 0.0150  & 1.412-1.418 \\ \hline
N(3)-N(3)                                                                                                                                            & -0.1170 & 0.1075  & 0.2943  & 1.384-1.457 \\ \hline
N(3)-N(2)                                                                                                                                            & 0.0426  & -0.1165 & -0.0721 & 1.345-1.375 \\ \hline
N(2)-N(2)(aromatic)                                                                                                                                  & 0.2894  & -0.1551 & -0.1649 & 1.287-1.375 \\ \hline
N(2)-N(2)                                                                                                                                            & 0.0477  & 0.0710  & -0.0100 & 1.202-1.262 \\ \hline
N(2)-N(1) (azides)                                                                                                                                   & 0.0000  & 0.0203  & 0.0003  & 1.114-1.137 \\ \hline
                                                                                                                                                    \multicolumn{5}{|c|}{Nitrogen-Oxygen bonds}             \\ \hline
N(3)-O(2)                                                                                                                                            & 0.5630  & 0.1865  & 0.5170  & 1.388-1.468 \\ \hline
N(3)-O(1)                                                                                                                                            & -0.1038 & -0.0670 & -0.0907 & 1.228-1.316 \\ \hline
N(2)-O(2)                                                                                                                                            & 0.6880  & 0.0412  & 0.8593  & 1.365-1.420 \\ \hline
N(3)-O(1)                                                                                                                                            & -0.0948 & 0.0117  & -0.0875 & 1.203-1.251 \\ \hline

\end{longtable}

\begin{longtable}[c]{|l|c|c|c|c|}
\caption{Values of the Kullback-Leibler (KL) divergence of the
  different training sets and the tautobase at specific intervals for
  different types of bonds for molecules with $n_{\rm{atoms}} >
  9$. The values of the bond lengths were taken from Allen, F.H.,
  \textit{et al.}, 2006 \cite{allen2006typical} }
\label{sitab:KL_9p}
\\\hline
Type of Bond                                                                                                                                         & QM9     & PC9     & ANI-1E    & Interval(\r{A})  \\ \hline
\endfirsthead
\caption{Values of the Kullback-Leibler (KL) divergence of the
  different training sets and the tautobase at specific intervals for
  different types of bonds for molecules with $n_{\rm{atoms}} >
  9$. The values of the bond lengths were taken from Allen, F.H.,
  \textit{et al.}, 2006 \cite{allen2006typical}(cont.) } \\
\hline
Type of Bond                                                                                                                                         & QM9     & PC9     & ANI-1E    & Interval(\r{A})  \\ \hline
\endhead
\hline \multicolumn{4}{r}{\textit{Continued on next page..}} \\
\endfoot
\hline
\endlastfoot
C(sp)-C(sp)(triple)                                                                                                                                  & 0.003   & 0.000   & 0.005   & 1.167-1.197 \\ \hline
C(sp$^2$)-C(sp$^2$)(double)                                                                                                                                & -0.121  & -0.022  & 0.000   & 1.280-1.405 \\ \hline
C(sp$^2$)-C(sp$^2$)(single)                                                                                                                                & 0.767   & 0.719   & 0.657   & 1.400-1.568 \\ \hline
C(sp$^3$)-C(sp$^3$)(single)                                                                                                                                & 0.937   & 0.853   & 0.746   & 1.458-1.610 \\ \hline
C(sp)-C(sp)(single)                                                                                                                                  & -0.183  & -0.217  & -0.158  & 1.374-1.474 \\ \hline
C(ar)-C(ar)                                                                                                                                          & -0.202  & -0.208  & -0.191  & 1.350-1.440 \\ \hline
C(sp$^3$)-C(sp$^2$)                                                                                                                                        & 0.499   & 0.482   & 0.582   & 1.470-1.538 \\ \hline
C(sp$^3$)-C(ar)                                                                                                                                         & 0.513   & 0.513   & 0.584   & 1.479-1.539 \\ \hline
C(sp$^3$)-C(sp)                                                                                                                                         & -0.014  & -0.031  & 0.003   & 1.436-1.481 \\ \hline
C(sp$^2$)-C(ar)                                                                                                                                         & 0.146   & 0.069   & 0.249   & 1.441-1.512 \\ \hline
C(sp$^2$)-C(sp)                                                                                                                                         & -0.021  & -0.024  & -0.018  & 1.425-1.441 \\ \hline
C(ar)-C(sp)                                                                                                                                          & -0.019  & -0.022  & -0.016  & 1.430-1.448 \\ \hline
                                                                                                                                                 \multicolumn{5}{|c|}{Carbon-Nitrogen bonds}   \\ \hline
C(sp$^3$)-N(4)                                                                                                                                          & 0.1634  & 0.0808  & 0.0236  & 1.482-1.510 \\ \hline
C(sp$^3$)-N(3)                                                                                                                                          & 0.5185  & 0.6255  & 0.4057  & 1.446-1.572 \\ \hline
C(sp$^3$)-N(2)                                                                                                                                          & 0.3322  & 0.4451  & 0.2122  & 1.461-1.506 \\ \hline
C(sp$^2$)-N(3)                                                                                                                                          & -0.2625 & -0.2728 & -0.2479 & 1.314-1.419 \\ \hline
C(sp$^2$)-N(2) (Imidazole)                                                                                                                              & -0.0560 & -0.0593 & -0.0518 & 1.369-1.384 \\ \hline
C(ar)-N(4)                                                                                                                                           & 0.0733  & 0.1890  & 0.1165  & 1.461-1.470 \\ \hline
C(ar)-N(3)                                                                                                                                           & -0.0104 & 0.2425  & 0.1849  & 1.340-1.476 \\ \hline
C(ar)-N(2)                                                                                                                                           & -0.0027 & -0.0028 & 0.0118  & 1.422-1.442 \\ \hline
C(sp$^2$)-N(3) (furoxan)                                                                                                                                & -0.0226 & -0.0240 & -0.0221 & 1.311-1.324 \\ \hline
C(sp$^2$)-N(2)                                                                                                                                          & -0.0917 & -0.0424 & -0.0127 & 1.273-1.339 \\ \hline
C(ar)-N(3)                                                                                                                                           & -0.0984 & -0.0977 & -0.1007 & 1.325-1.369 \\ \hline
C(ar)-N(2)                                                                                                                                           & -0.0855 & -0.0850 & -0.0863 & 1.300-1.348 \\ \hline
C(sp)-N(2)                                                                                                                                           & 0.0136  & 0.0005  & 0.0028  & 1.140-1.148 \\ \hline
C(sp)-N(1)                                                                                                                                           & -0.1404 & -0.1741 & -0.0339 & 1.131-1.449 \\ \hline
                                                                                                                                                 \multicolumn{5}{|c|}{Carbon-Oxygen bonds}    \\ \hline
C(sp$^3$)-O(2) (Alcohols)                                                                                                                               & 0.6515  & 0.5870  & 0.5996  & 1.395-1.449 \\ \hline
C(sp$^3$)-O(2)(Dialkyl ethers)                                                                                                                          & 0.6108  & 0.5655  & 0.5093  & 1.405-1.458 \\ \hline
C(sp$^3$)-O(2)(aryl alkyl ethers)                                                                                                                       & 0.3302  & 0.3556  & 0.2713  & 1.417-1.438 \\ \hline
C(sp$^2$)-O(2)\footnote{Aryl alkyl ethers, alkyl   esters of carboxilic acids, alkyl esters of alpha, beta unsaturated acids,   alkyl esterets of benzoic acid} & 0.1719  & 0.0874  & 0.0270  & 1.435-1.501 \\ \hline
C(sp$^2$)-O(2)(Ring systems)                                                                                                                            & 0.2398  & 0.1593  & 0.0673  & 1.430-1.501 \\ \hline
C(sp$^2$)-O(2)(Enols)                                                                                                                                   & -0.0349 & -0.0245 & -0.0335 & 1.324-1.342 \\ \hline
C(sp$^2$)-O(2)(enol esters)                                                                                                                             & -0.0418 & -0.0013 & -0.0211 & 1.341-1.363 \\ \hline
C(sp$^2$)-O(2)(acids)                                                                                                                                   & -0.0261 & -0.0196 & -0.0271 & 1.279-1.320 \\ \hline
C(sp$^2$)-O(2)(esters)                                                                                                                                  & 0.2461  & 0.2170  & 0.3624  & 1.328-1.420 \\ \hline
C(sp$^2$)-O(2)(anhydrides)                                                                                                                              & 0.0175  & -0.0017 & 0.0245  & 1.379-1.393 \\ \hline
C(sp$^2$)-O(2)(ring systems)                                                                                                                            & -0.0718 & -0.0090 & -0.0364 & 1.332-1.377 \\ \hline
C(ar)-O(2)(Phenols)                                                                                                                                  & -0.0261 & 0.0091  & -0.0040 & 1.353-1.373 \\ \hline
C(ar)-O(2)(aryl alkyl ethers)                                                                                                                        & -0.0112 & 0.0045  & 0.0018  & 1.363-1.377 \\ \hline
C(ar)-O(2)(diaryl ethers)                                                                                                                            & 0.0114  & -0.0029 & 0.0192  & 1.375-1.391 \\ \hline
C(ar)-O(2)(esteres)                                                                                                                                  & 0.1064  & 0.0513  & 0.1394  & 1.394-1.408 \\ \hline
C(sp$^2$)-O(1)(double) Aldehydes and   ketones)                                                                                                         & -0.1140 & -0.1154 & -0.1149 & 1.188-1.238 \\ \hline
C(sp$^2$)-O(1)(double)\footnote{Delocalized double   bonds in carboylate anions}                                                                               & -0.0404 & -0.0553 & -0.0426 & 1.232-1.262 \\ \hline
C(sp$^2$)-O(1)(double)(Carboxylic acids)                                                                                                                & -0.0978 & -0.0981 & -0.0938 & 1.203-1.241 \\ \hline
C(sp$^2$)-O(1)(double)(esters)                                                                                                                          & -0.0373 & -0.0375 & -0.0425 & 1.181-1.207 \\ \hline
C(sp$^2$)-O(1)(anhydrides)                                                                                                                              & -0.0143 & -0.0118 & -0.0150 & 1.184-1.193 \\ \hline
C(sp$^2$)-O(1)(lactones)                                                                                                                                & 0.0000  & 0.0000  & 0.0000  & 1.187-1.209 \\ \hline
C(sp$^2$)-O(1)(double)(amides)                                                                                                                          & -0.1143 & -0.1195 & -0.1149 & 1.193-1.243 \\ \hline
                                                                                                                                                    \multicolumn{5}{|c|}{Nitrogen-Nitrogen bonds}     \\ \hline
N(4)-N(3)                                                                                                                                            & -0.0088 & 0.0221  & 0.0297  & 1.412-1.418 \\ \hline
N(3)-N(3)                                                                                                                                            & -0.0963 & 0.2138  & 0.4141  & 1.384-1.457 \\ \hline
N(3)-N(2)                                                                                                                                            & 0.0599  & -0.1094 & -0.0745 & 1.345-1.375 \\ \hline
N(2)-N(2)(aromatic)                                                                                                                                  & 0.2938  & -0.1645 & -0.1646 & 1.287-1.375 \\ \hline
N(2)-N(2)                                                                                                                                            & 0.0053  & 0.0423  & -0.0506 & 1.202-1.262 \\ \hline
N(2)-N(1) (azides)                                                                                                                                   & 0.0000  & 0.0203  & 0.0003  & 1.114-1.137 \\ \hline
                                                                                                                                                 \multicolumn{5}{|c|}{Nitrogen-Oxygen bonds}     \\ \hline
N(3)-O(2)                                                                                                                                            & 0.8954  & 0.4577  & 0.8386  & 1.388-1.468 \\ \hline
N(3)-O(1)                                                                                                                                            & -0.1448 & -0.1724 & -0.1143 & 1.228-1.316 \\ \hline
N(2)-O(2)                                                                                                                                            & 1.0017  & 0.1725  & 1.2526  & 1.365-1.420 \\ \hline
N(3)-O(1)                                                                                                                                            & -0.1417 & -0.1208 & -0.1158 & 1.203-1.251 \\ \hline

\end{longtable}

\begin{table}[hbtp]
\caption{Mean Absolute (MAE) and Root-Mean-Squared Error (RMSE) for
  the prediction of tautomerization energy $\Delta E_{\rm Tauto}$, and
  the single isomer energies, $E_{\rm SI}$, for the entire Tautobase
  (1257 tautomeric pairs) for each of the datasets evaluated for the
  MMFF94 force field geometries.}
\begin{tabular}{l|cc|cc|}
\toprule
 & \multicolumn{2}{c}{$ \Delta E_{\textrm{Tauto}}$} & \multicolumn{2}{c}{$E_{\textrm{SI}}$}        \\ 
Database & MAE                 & RMSE               & MAE           & RMSE  \\ \midrule
QM9      & 7.40                & 10.60               & 8.36       &  11.76 \\ 
PC9      & 6.59                & 9.72               & 9.65       &  14.95 \\ 
ANI-1E    & 6.11                & 9.31               & 16.70      & 16.70 \\ 
ANI-1    & 5.57                & 8.79               & 20.17      & 30.35 \\ 
ANI-1x   & 3.42                  & 5.79             & 5.97       &  8.38  \\ \bottomrule
\end{tabular}
\label{sitab:mae_MMFF94}
\end{table}

\begin{figure}
    \centering
    \includegraphics[width=0.85\textwidth]{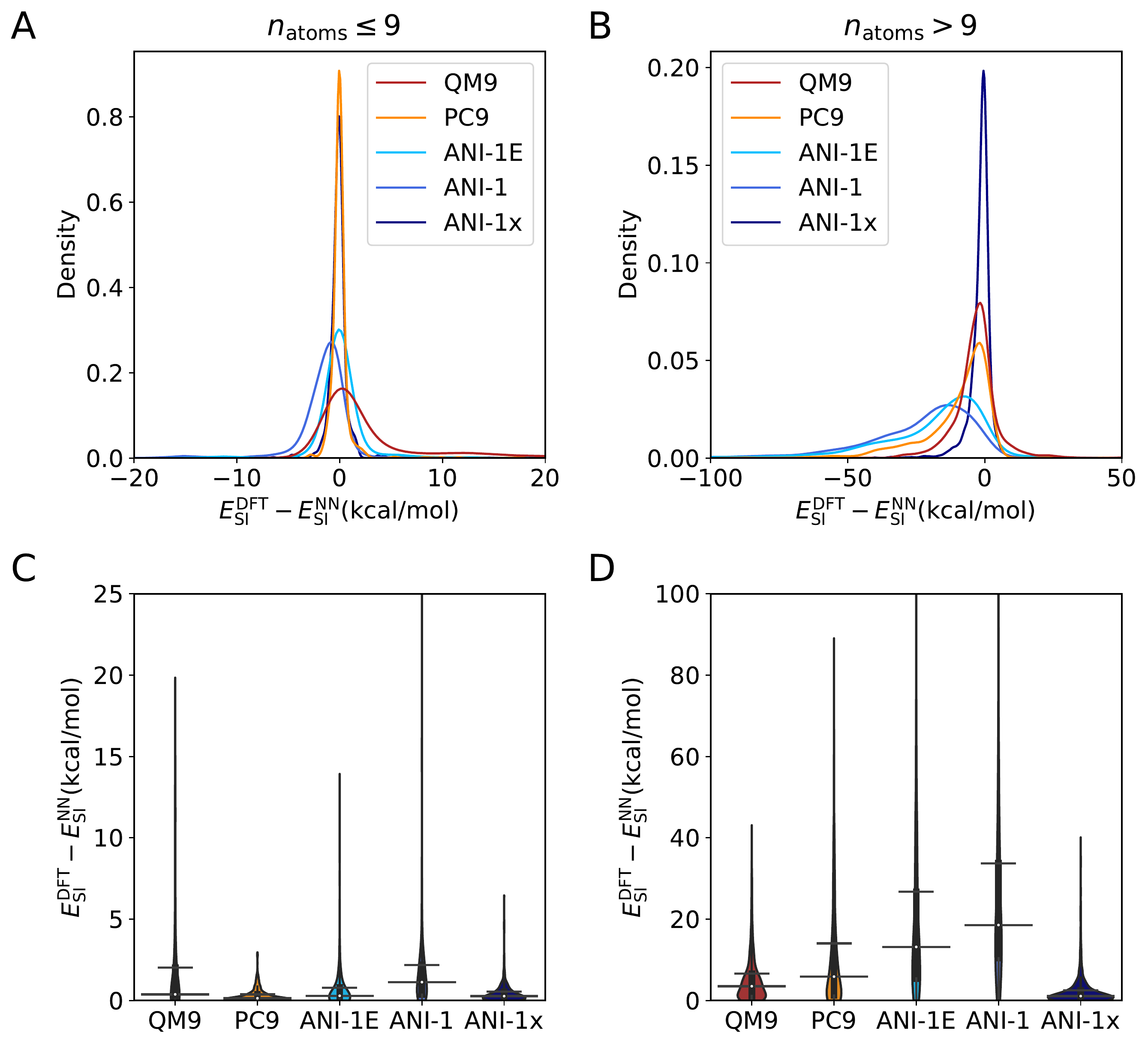}
    \caption{Error analysis on the prediction of energies of the
      single molecules. Panels A and B: Kernel density estimate for
      prediction of the energies of single isomer for the different
      databases. Panels C and D: Normalized error distribution up to
      the 95\% quantile of the different datasets evaluated on this
      work for the energy of a single isomer.  The blackbox inside spans
      between the 25\% and 75\% quantiles with a white dot indicating
      the mean of the distribution. The whisker marks indicate the 5\%
      and 95 \% quantiles. The left and right columns are for Set1 and
      Set2, respectively.}
    \label{sifig:error_dist}
\end{figure}

\begin{figure}[h!]
    \centering
    \includegraphics[scale=0.47]{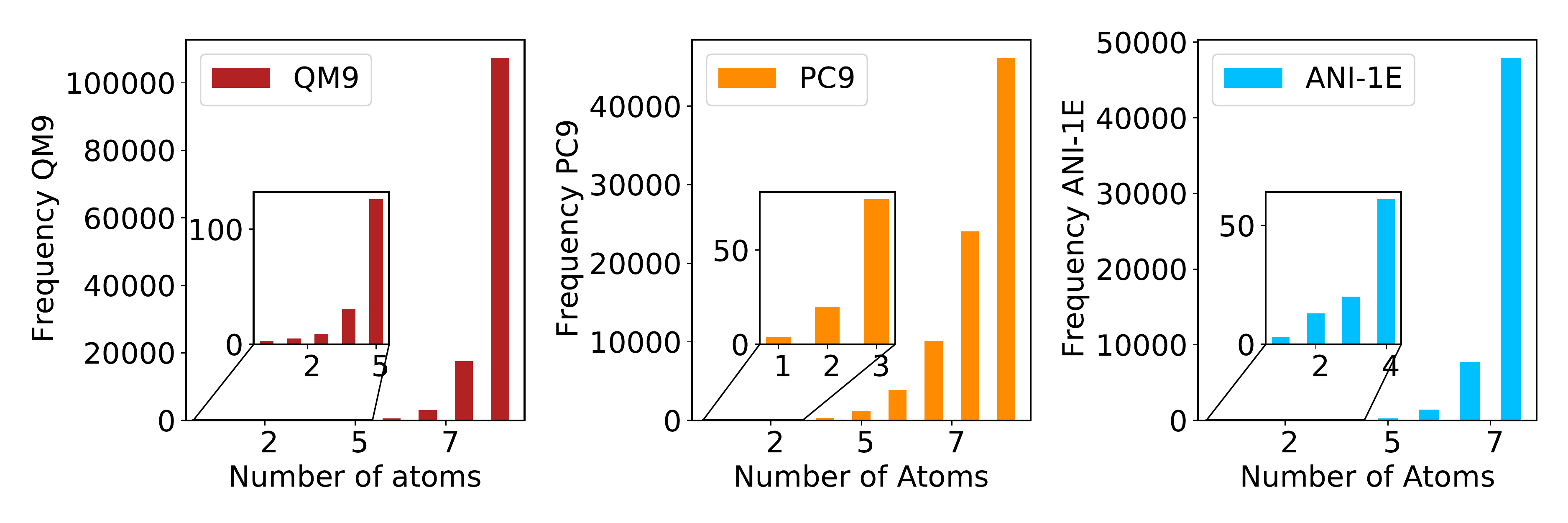}
    \caption{Number of heavy atoms (C,N,O) in the QM9, PC9, and ANI-1E
      databases (from left to right) used in the present work. The
      insets show enlargements for cases with few representatives.}
    \label{sifig:NatomsperDatabases}
\end{figure}

\begin{figure}[h!]
    \centering
    \includegraphics[scale=0.45]{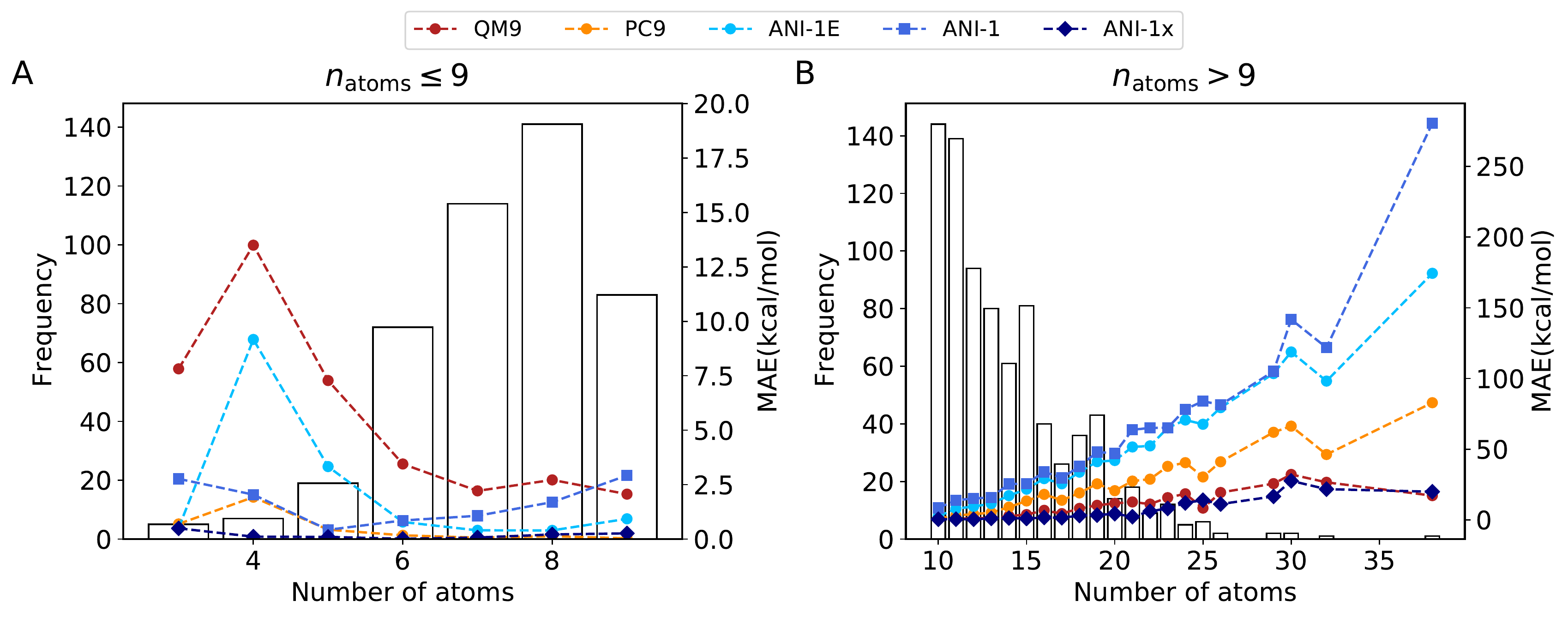}
    \caption{Mean Absolute Error (MAE) by number of heavy atoms
      (C,O,N) for $E_{\rm SI}$. A histogram of the number of molecules
      for different numbers of heavy atoms is shown in the
      background. Panel A for Set1 and panel B for Set2.}
    \label{sifig:MAEvNAtom_mol}
\end{figure}

\begin{figure}[h!]
    \centering
    \includegraphics[scale=0.5]{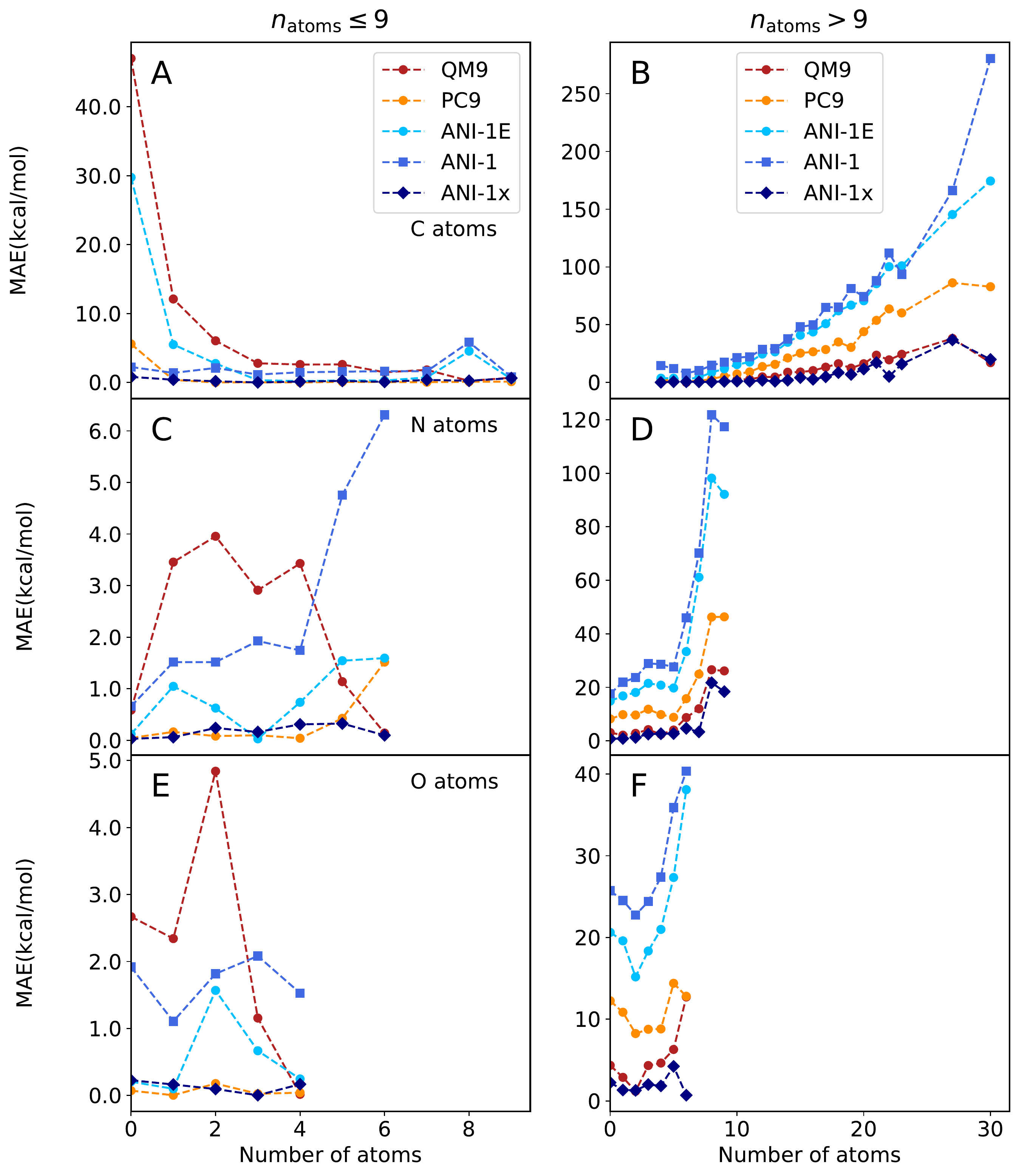}
    \caption{Mean Absolute Error (MAE) by number of atoms of a given
      element (carbon (top), nitrogen (middle), oxygen (bottom)) for
      the energy of a single isomer $E_{\rm SI}$. Panels A and B shows
      the results by number of C atoms. Panels C and D shows the
      results by number of N atoms. Finally, panels E and F show the
      results by number of O atoms. Left and right columns for Set1
      and Set2, respectively.}
       \label{sifig:natomselementvsmae}
\end{figure}

\begin{figure}[h!]
    \centering \includegraphics[scale=0.45]{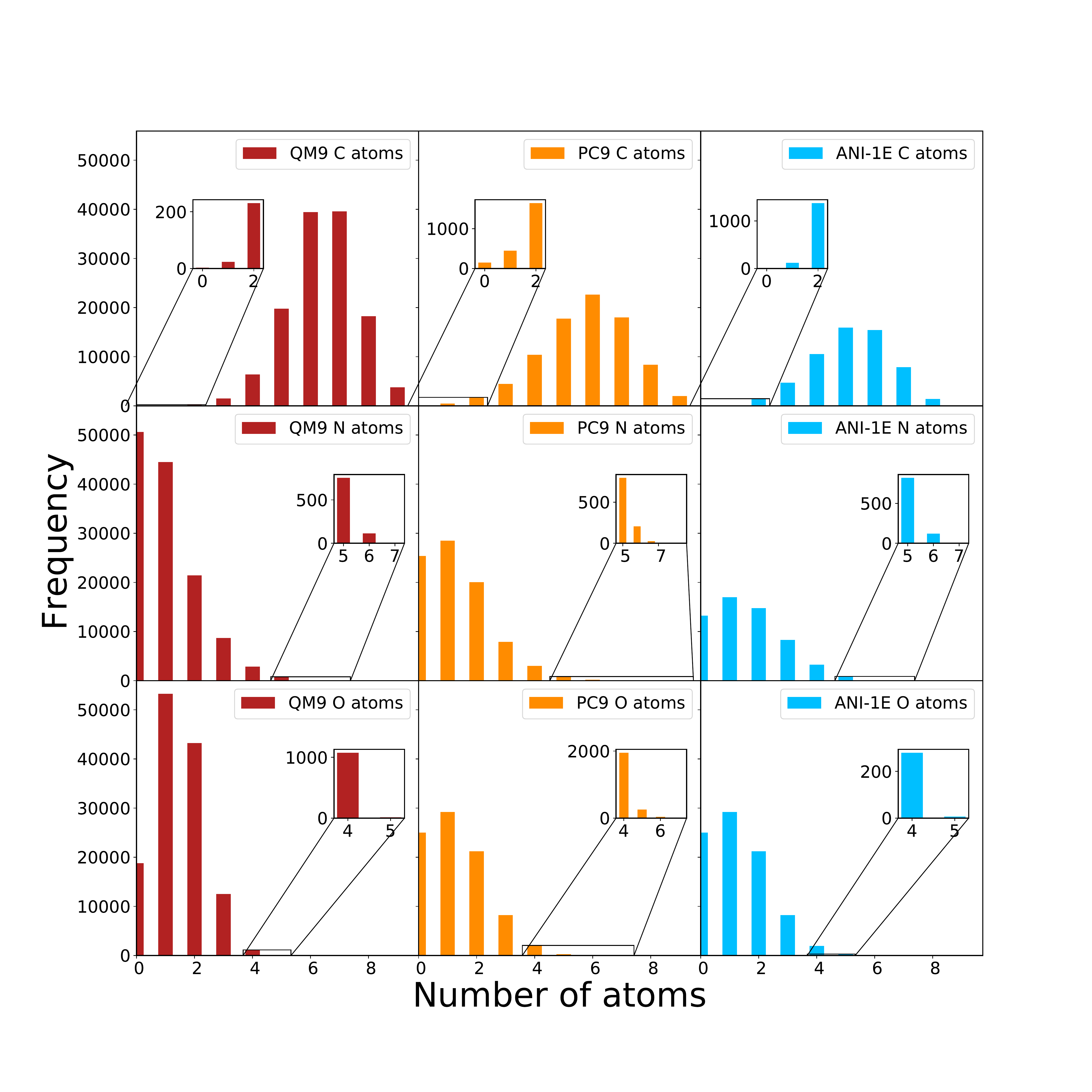}
    \caption{Number of C- ,N-, and O- atoms (from top to bottom) in
      the QM9, PC9, and ANI-1E databases (from left to right) used in
      the present work. The insets show enlargements for cases with
      few representatives.}
    \label{sifig:NatomselementperDatabases}
\end{figure}

\begin{figure}[h!]
    \centering
    \includegraphics[scale=0.5]{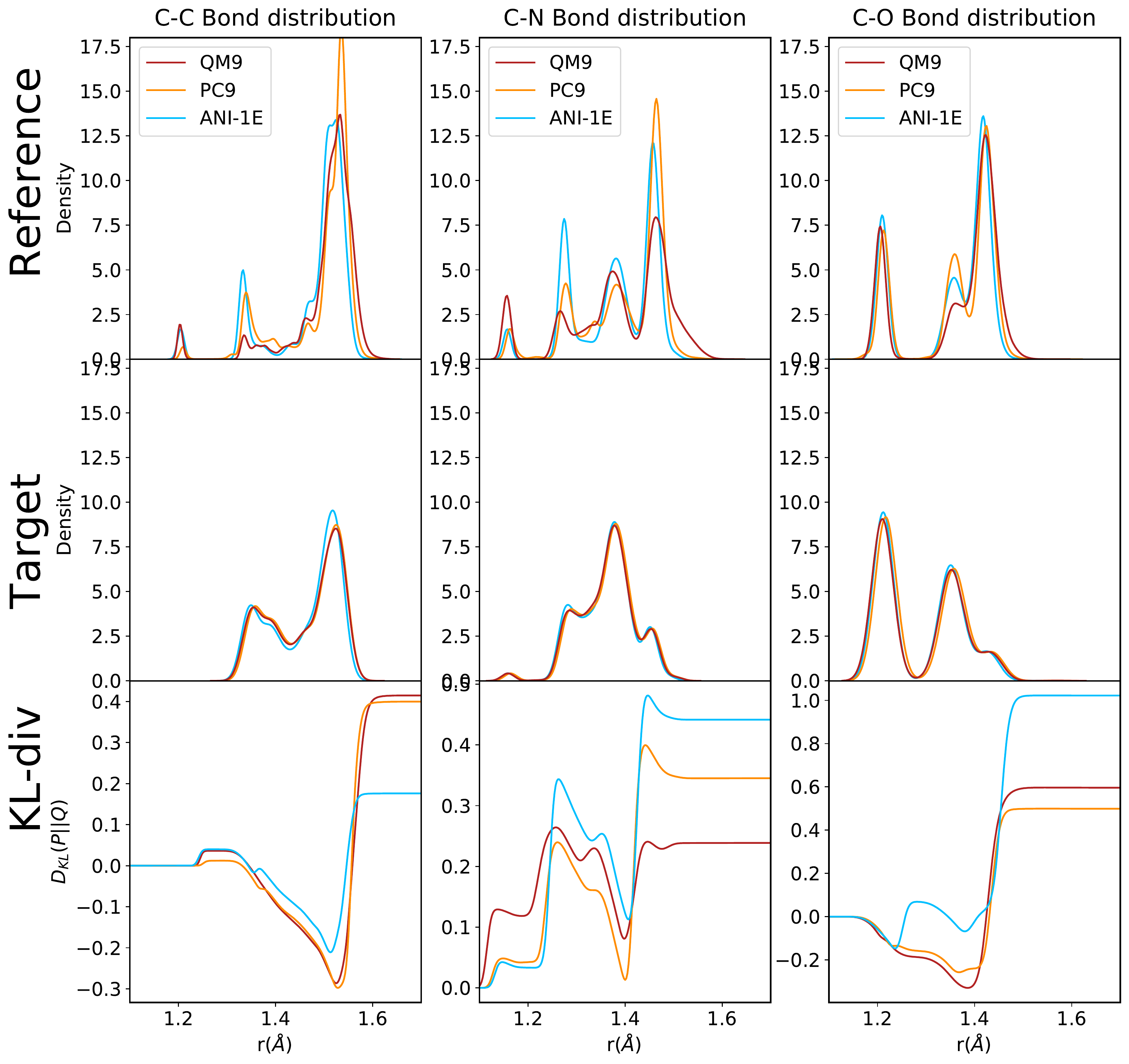}
    \caption{Kernel distributions of different types of bond lengths
      involving C atoms. Top row are the results for the reference
      sets (PC9, QM9 and ANI-1E). Middle row shows the results of the
      geometries of tautobase optimized at level of theory of the
      different databases for Set1. The KL-divergence between
      reference and target data set distributions is reported in the
      bottom row.}
    \label{sifig:dist_bonds_C_9}
\end{figure}

\begin{figure}[h!]
    \centering
    \includegraphics[scale=0.5]{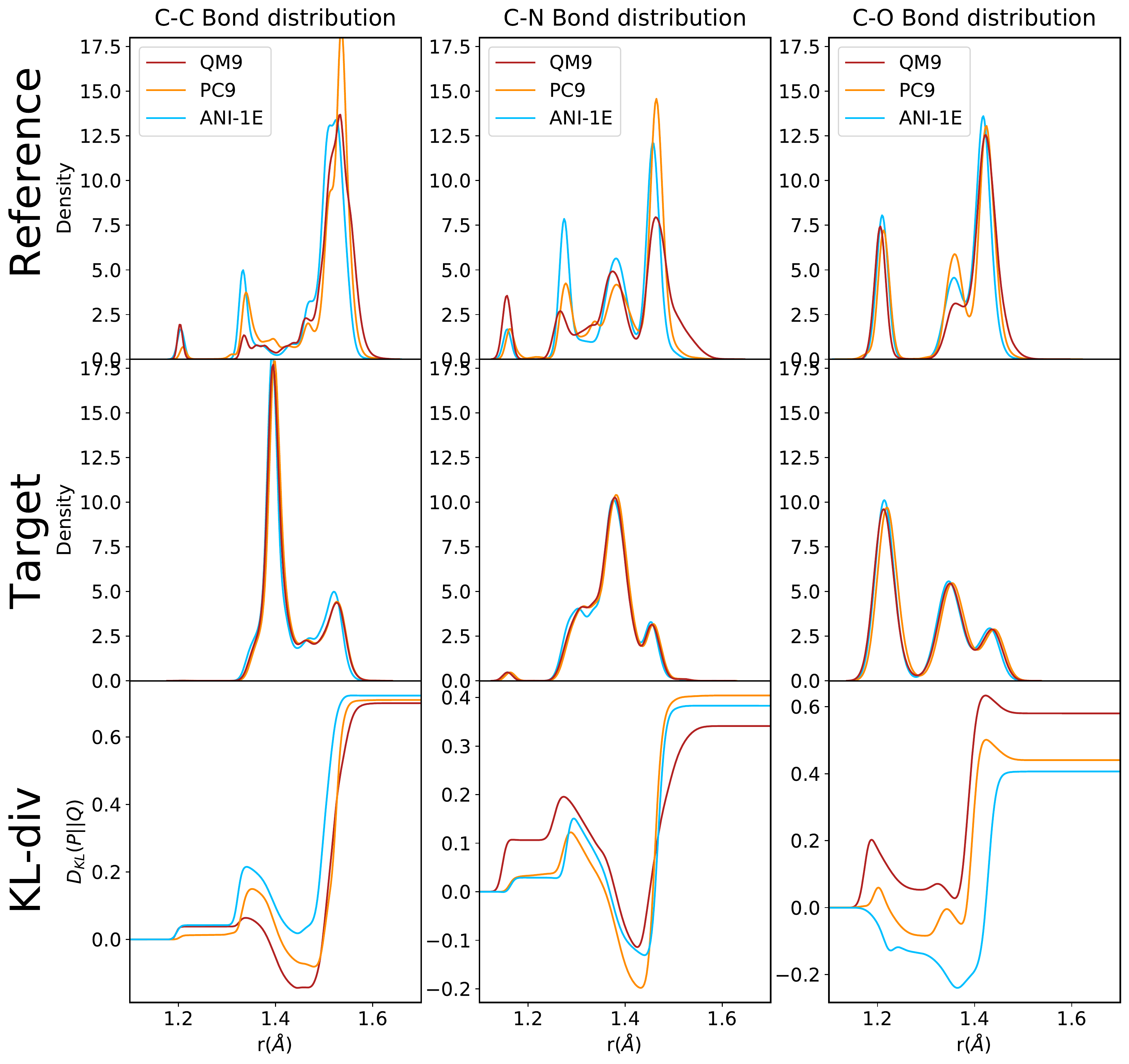}
    \caption{Kernel distributions of different types of bond lengths
      involving C atoms. Top row are the results for the reference
      sets (PC9, QM9 and ANI-1E). Middle row shows the results of the
      geometries of tautobase optimized at level of theory of the
      different databases for Set2. The KL-divergence between
      reference and target data set distributions is reported in the
      bottom row.}
    \label{sifig:dist_bonds_C_9p}
\end{figure}

\begin{figure}[h!]
    \centering
    \includegraphics[scale=0.5]{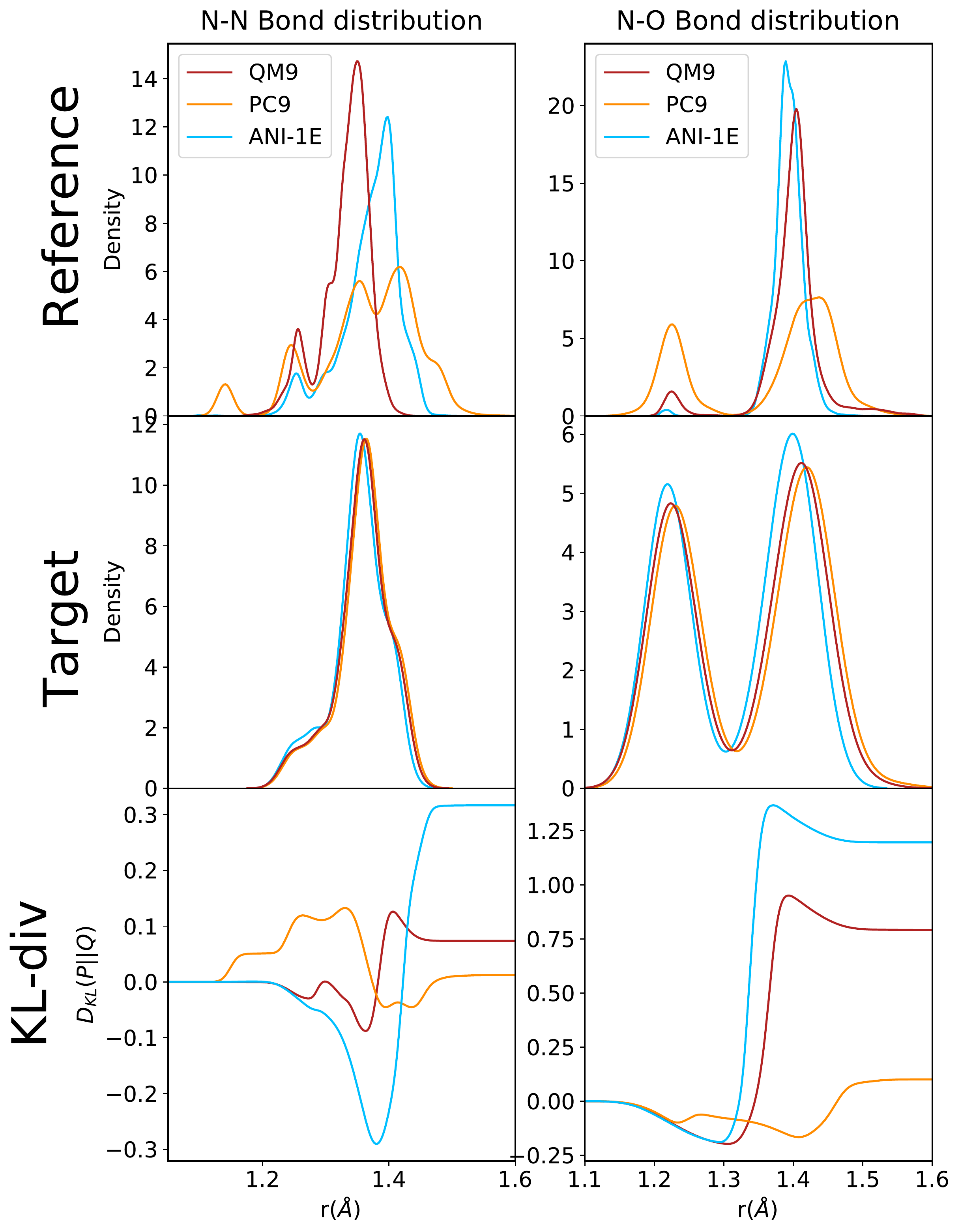}
    \caption{Kernel distributions of different types of bond lengths
      involving N atoms. Top row are the results for the reference
      sets (PC9, QM9 and ANI-1E). Middle row shows the results of the
      geometries of tautobase optimized at level of theory of the
      different databases for Set1. The KL-divergence between
      reference and target data set distributions is reported in the
      bottom row..}
    \label{sifig:dist_bonds_N_9}
\end{figure}

\begin{figure}[h!]
    \centering
    \includegraphics[scale=0.5]{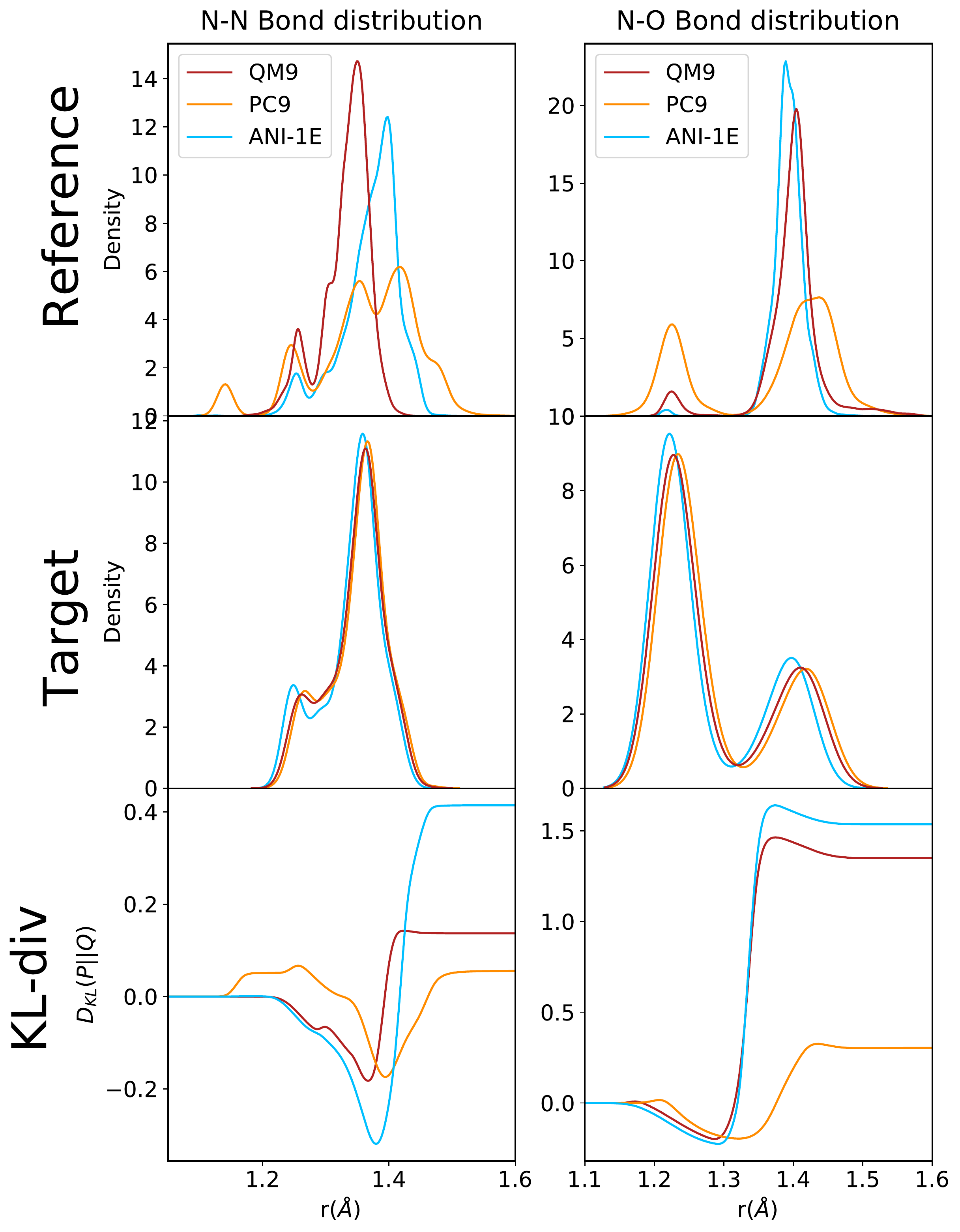}
    \caption{Kernel distributions of different types of bond lengths
      involving C atoms. Top row are the results for the reference
      sets (PC9, QM9 and ANI-1E). Middle row shows the results of the
      geometries of tautobase optimized at level of theory of the
      different databases for Set2. The KL-divergence between
      reference and target data set distributions is reported in the
      bottom row.}
    \label{sifig:dist_bonds_N_9p}
\end{figure}


\begin{figure}
    \centering
    \includegraphics[width=0.85\textwidth]{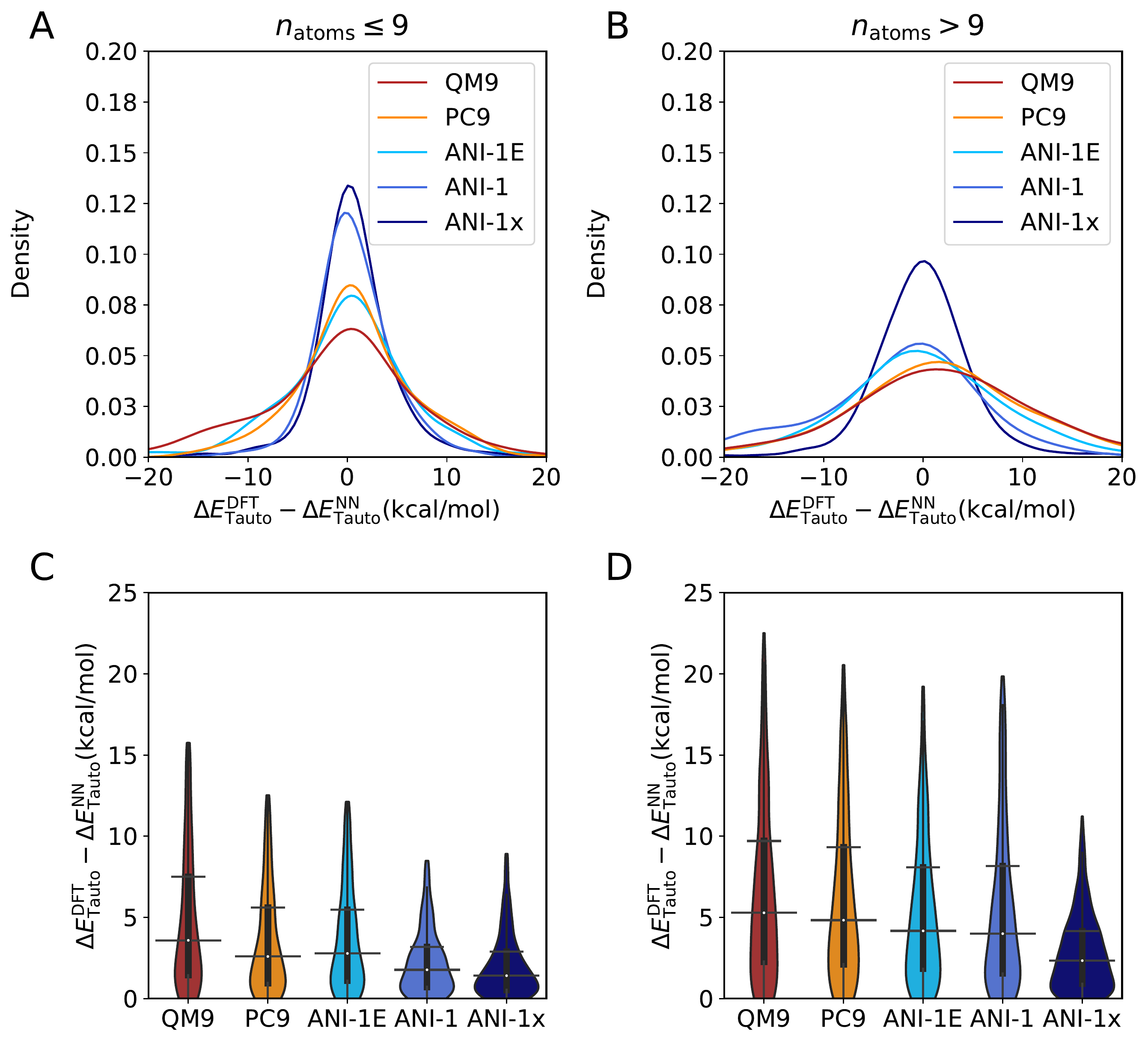}
    \caption{Error analysis on the prediction of the tautomerization
      energies using an optimized geometry with MMFF94 force
      field. Panels A and B: Kernel density estimate for prediction of
      the tautomerization energies for the different databases. Panels
      C and D: Normalized error distribution up to the 95\% quantile
      of the different datasets evaluated on this work for the
      tautomerization energy.  The blackbox inside spans between the
      25\% and 75\% quantiles with a white dot indicating the mean of
      the distribution. The whisker marks indicate the 5\% and 95 \%
      quantiles. The left and right columns are for Set1 and Set2,
      respectively}
    \label{sifig:error_dist_MMFF94}
\end{figure}

\begin{figure}[h!]
    \centering
    \includegraphics[scale=0.45]{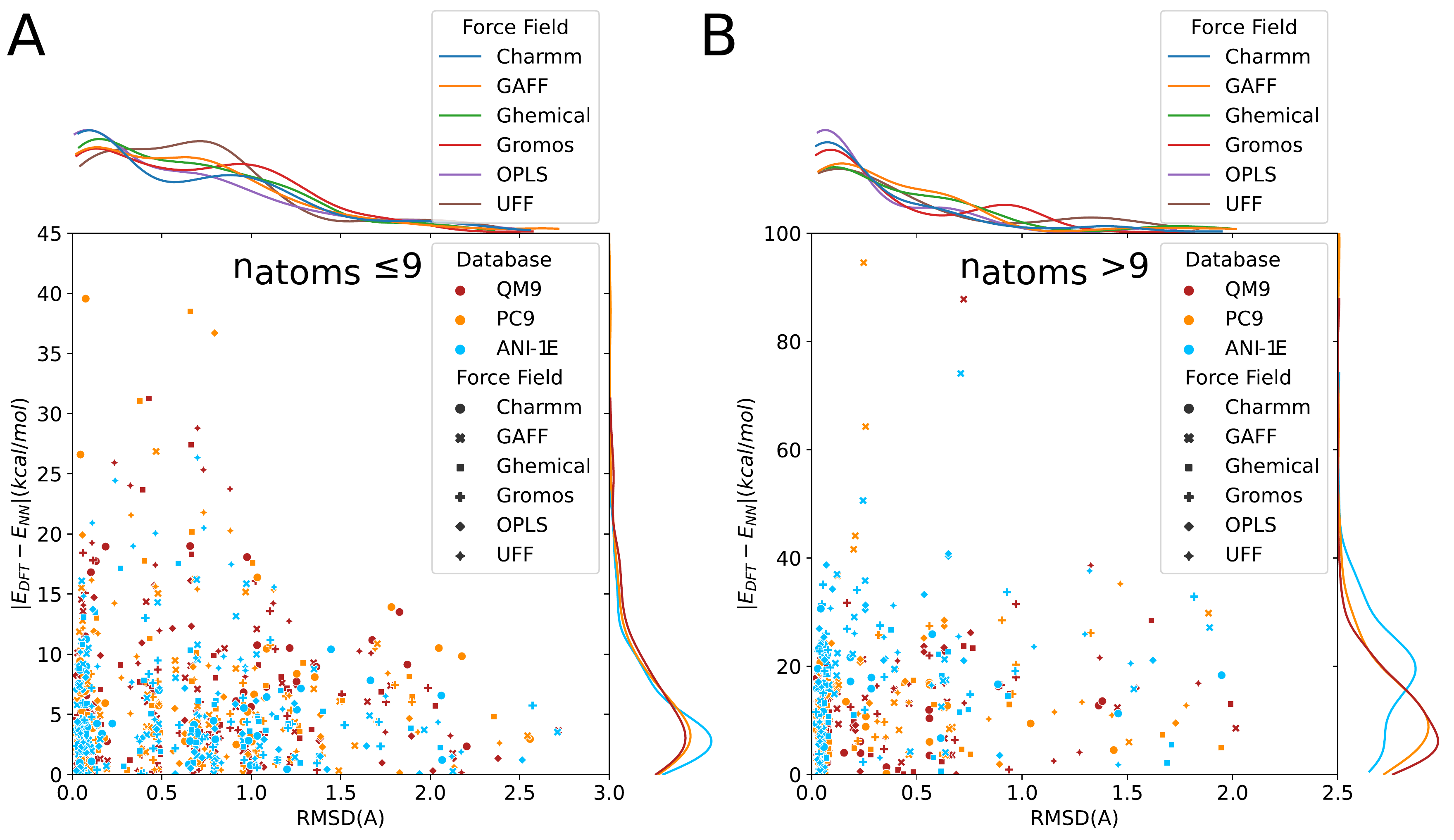}
    \caption{Effect of the initial geometry on the prediction of the
      energy by the NN models for the molecules from the SAMPL2
      challenge\cite{geballe2010sampl2}. Plots of the Root Mean Square
      Displacement with respect to the optimized geometry vs.\ the
      absolute error between the energy obtained from DFT calculations
      and the energy from the NN model for the molecules on panel A
      results for isomers with $n_{\rm{atoms}} \leq 9$ and on panel B
      for$n_{\rm{atoms}} > 9$. In both plots, the X axis shows the
      kernel distribution function for the RMSD for the different
      force fields used to generate the test geometries. In the Y axis
      is shown the Kernel distribution function for the predicted
      energies with the different databases.}
    \label{sifig:RMSD_scatter}
\end{figure}

\begin{figure}[h!]
    \centering
\begin{subfigure}{.45\textwidth}
  \centering
  \includegraphics[width=.98\linewidth]{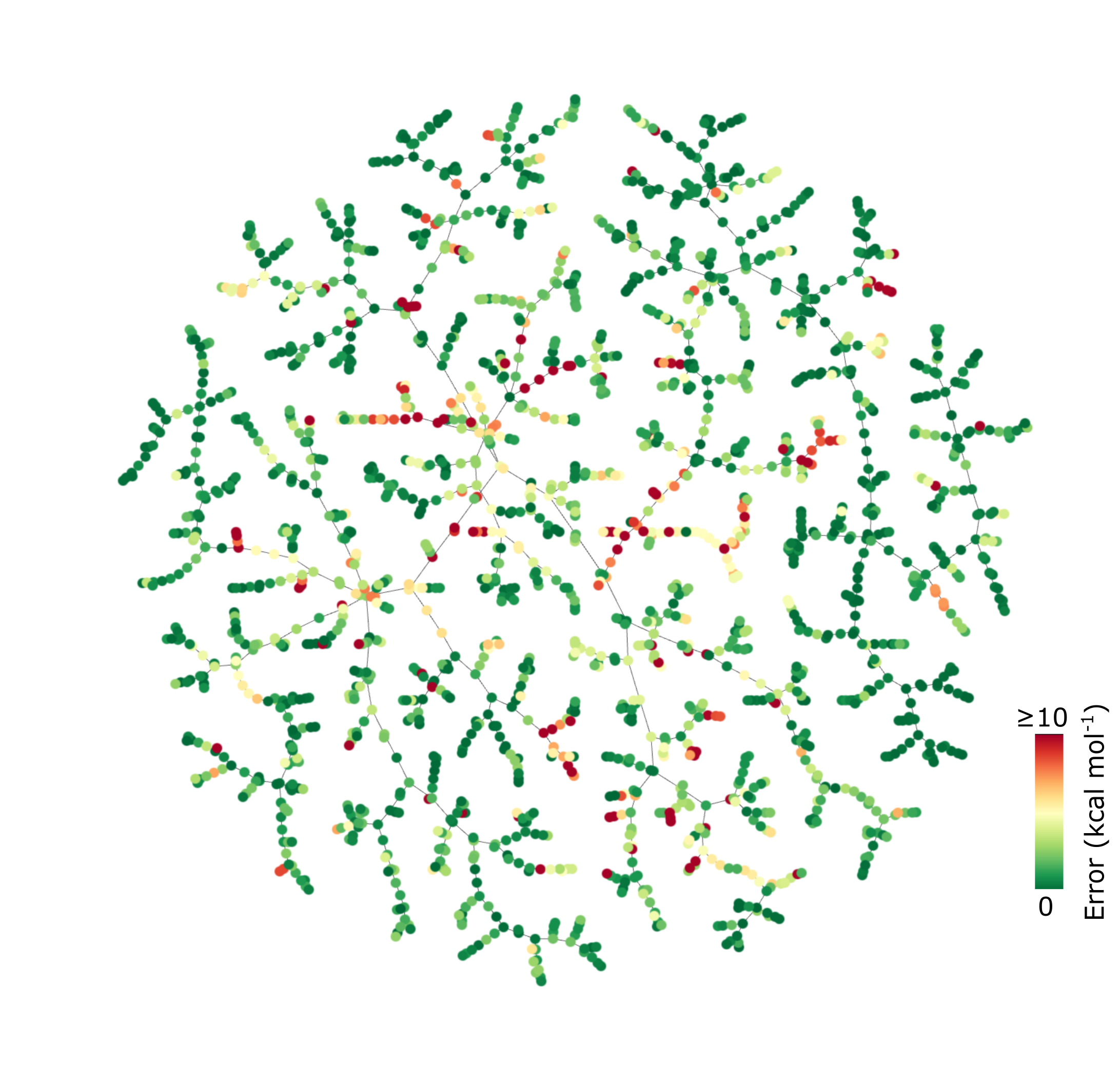}
  \caption{PC9}
  \label{sifig:sfig1}
\end{subfigure}    
\begin{subfigure}{.45\textwidth}
  \centering
  \includegraphics[width=.98\linewidth]{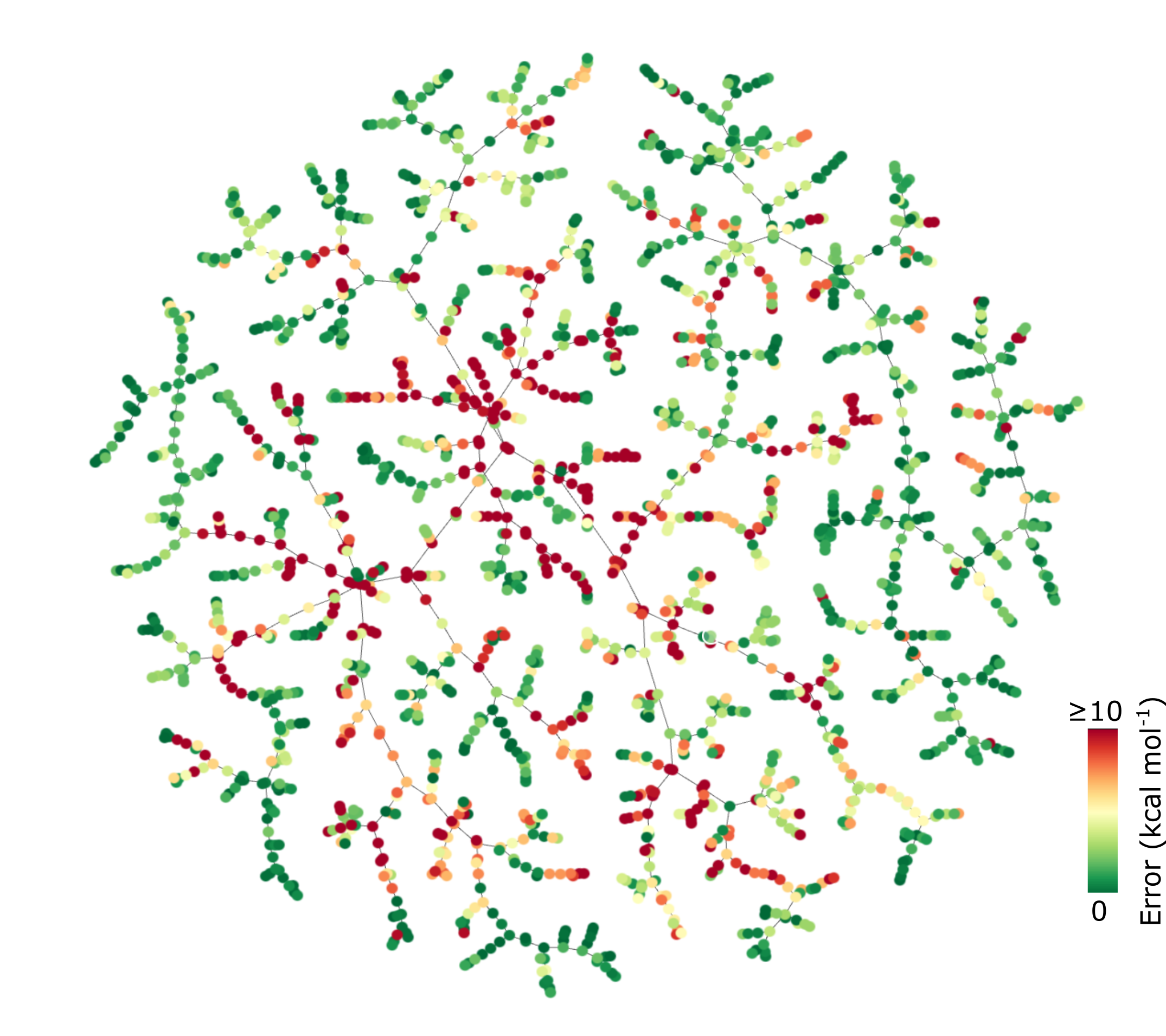}
  \caption{ANI-1}
  \label{sifig:sfig2}
\end{subfigure}%
\begin{subfigure}{.45\textwidth}
  \centering
  \includegraphics[width=.98\linewidth]{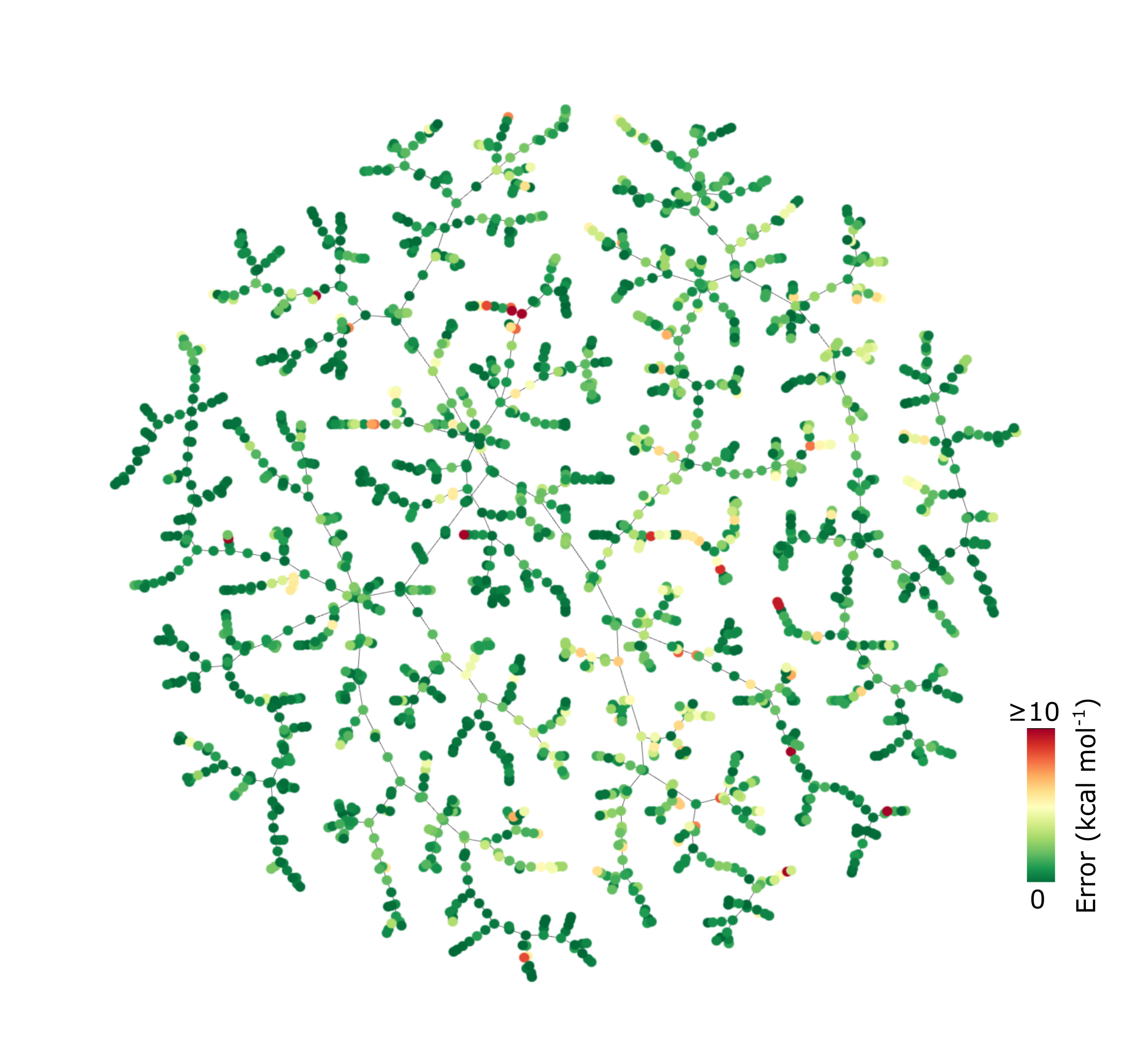}
  \caption{ANI-1x}
  \label{sifig:sfig3}
\end{subfigure}
\begin{subfigure}{.45\textwidth}
  \centering
  \includegraphics[width=.98\linewidth]{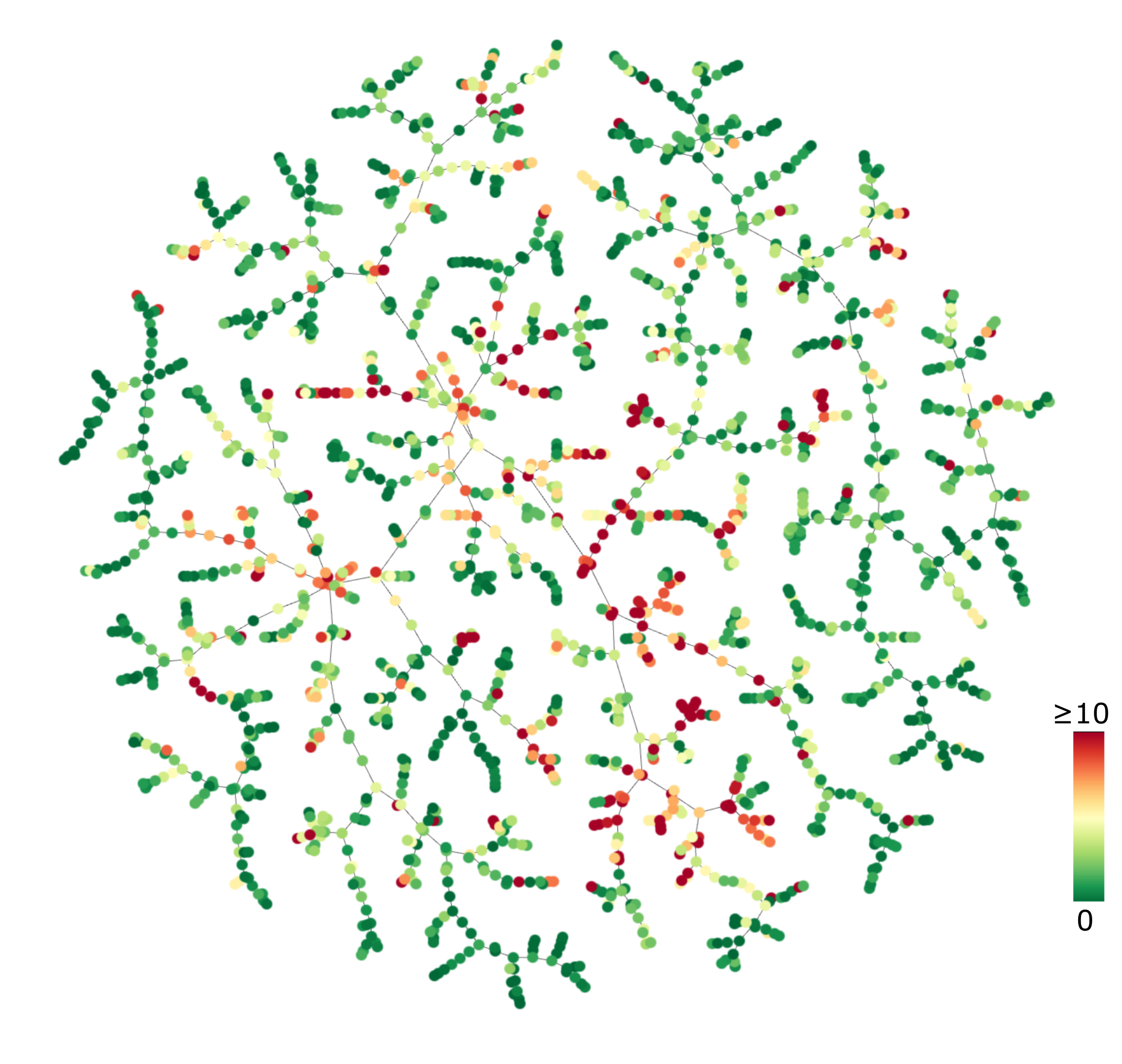}
  \caption{ANI-1E}
  \label{sifig:sfig4}
\end{subfigure}
    \caption{TMAP projection of chemical space for all molecules in
      the TautoBase, coloured by error in tautomerization energy
      calculated using the ML potentials trained on (a) PC9, (b)
      ANI-1, (c) ANI-1x and (d) ANI-1E. }
    \label{sifig:TMAP_SI}
\end{figure}

\clearpage
\bibliography{biblio1}